\documentclass[final,authoryear,5p,times,twocolumn]{elsarticle}

\usepackage{graphicx}
\usepackage{lscape}
\usepackage{setspace}
\usepackage{caption}
\usepackage{amssymb,amsmath}
\usepackage{lipsum}
\usepackage{multirow}
\usepackage{booktabs}
\usepackage{hhline}
\usepackage{tabularx}
\usepackage{dcolumn}
\usepackage{placeins}
\usepackage{float}
\usepackage{subfigure}
\restylefloat{table}
\usepackage{textcomp}
\usepackage{url}
\usepackage{pdfpages}

\journal{Icarus}

\begin{document}

\begin{frontmatter}


\author{Abedin Abedin\corref{cor1}\fnref{label1}}
\ead{aabedin@uwo.ca}
\author{Pavel Spurn\'y\fnref{label2}}
\author{Paul Wiegert\fnref{label1}}
\author{Petr Pokorn\'y\fnref{label1}}
\author{Ji\v{r}\'{\i} Borovi\v{c}ka\fnref{label2}}
\author{Peter Brown\fnref{label1}}

\cortext[cor1]{Abedin Abedin}
\address[label1]{Department of Physics and Astronomy, The University of Western
Ontario, London, Canada N6A 3K7}
\address[label2]{Astronomical Institute, Academy of Sciences of the Czech Republic, Ondrejov, CZ-25165 Czech Republic}
\fntext[label3]{}

\title{On the age and formation mechanism of the core of the Quadrantid meteoroid stream}




\begin{abstract}

\footnotesize{The Quadrantid meteor shower is among the strongest annual meteor showers, and has drawn the attention of scientists for several decades. The stream is unusual, among others, for several reasons: its very short duration around maximum activity ($\approx$ 12 - 14 hours) as detected by visual, photographic and radar observations, its recent onset (around 1835 AD \citep{quet}) and because it had been the only major stream without an obvious parent body until 2003. Ever since, there have been debates as to the age of the stream and the nature of its proposed parent body, asteroid 2003 EH$_1$. 

\indent In this work, we present results on the most probable age and formation mechanism of the narrow portion of the Quadrantid 
meteoroid stream. For the first time we use data on eight high precision photographic Quadrantids, equivalent to gram - kilogram 
size, to constrain the most likely age of the core of the stream. Out of eight high-precision photographic Quadrantids, five 
pertain directly to the narrow portion of the stream. In addition, we also use data on five high-precision radar Quadrantids, 
observed within the peak of the shower.

\indent We performed backwards numerical integrations of the equations of motion of a large number of 'clones' of both, the eight 
high-precision photographic and five radar Quadrantid meteors, along with the proposed parent body, 2003 EH$_1$. According to our 
results, from the backward integrations, the most likely age of the narrow structure of the Quadrantids is between 200 - 300 
years. These presumed ejection epochs, corresponding to 1700 - 1800 AD, are then used for forward integrations of large numbers 
of hypothetical meteoroids, ejected from the parent 2003 EH$_1$, until the present epoch. The aim is to constrain whether the core 
of the Quadrantid meteoroid stream is consistent with a previously proposed relatively young age ($\approx$ 200 years).} 

\end{abstract}


\end{frontmatter}


\section{Introduction}
\label{intro}

The Quadrantids are among the most active meteor showers, reaching a peak activity of Zenithal Hourly Rate (ZHR) $\sim$ 110 - 130 on 3-4 January each year \citep{shelton65, hind70, hugh77}, as determined by photographic, visual, video and radar techniques. The stream has recently been linked with asteroid 2003 EH$_1$ \citep{jen04}. 

\indent The Quadrantid shower is unusual among meteoroid streams presently visible at the Earth for several reasons. Firstly, the Quadrantid meteor shower has a short duration of maximum activity, which we will call the 'core' or the "narrow structure" of the stream. The Full-Width at Half-Maximum (FWHM) of the core activity is $\approx$ 0.6 days \citep{shelton65, hugh77, brower06} for visual-sized particles, implying that this central portion is very young, while the shower as a whole has an overall duration of significant length $\sim $ 4 days. Secondly, it has only become active recently, being first noted circa 1835 \citep{quet, fisher30}. Moreover, the activity of the shower has changed dramatically over the last 150 years, from a very weak shower to among the strongest visible at the Earth \citep{jen06}. Finally, recent radar observations \citep{brown10} suggest low level activity of the shower persisting for a few months (November to mid January), suggesting the stream has an older component as well.

\indent Presently, the presumed parent body of the core of the Quadrantids is the {\textit{Near Earth Object}} (NEO) 2003 EH$_1$. 
6 Throughout this paper we will refer to it as asteroid 2003 EH$_1$. The object has been classified as an Amor type asteroid, 
although its nature is arguable based on dynamical criteria. Asteroid 2003 EH$_1$ has a short-period comet-like orbit, with a 
Tisserand parameter with respect to Jupiter of $T_{J} \approx $ 2.0, but currently shows no evidence of cometary activity, 
suggesting that it is a strong candidate for either a recently dormant or extinct comet \citep{koten06}. 

\indent Prior to the discovery of 2003 EH$_1$, a few other objects with less similar orbits had been connected to the Quadrantid meteoroid stream, most notably comet 96P/Machholz \citep{jones93, mcintosh90, bab92} and comet C/1490 Y1 \citep{hasegawa79, Lee09, Williams93, Williams98}. However, the relationship of these bodies to the Quadrantids remains unclear.

\indent The earliest attempt to tackle the age of the Quadrantid meteoroids stream can be attributed to \citet{hamid63}. The authors carried out a numerical secular perturbation analysis on the orbit of six doubly photographed Quadrantids and discovered large variations in the perihelion distance and the inclination of the stream orbit, with a period of 4000 years. Based on the backward secular solutions, the authors argued that the orbital elements of the six meteors were similar around 3000 years ago.

\indent \citet{williams79} calculated the secular variations of the orbital elements of the mean Quadrantid stream and concluded that the Quadrantid meteoroid stream may have resulted from two major comet break-ups about 1690 and 1300 years in the past, where the resulting meteoroids converged into their present orbit around 200 - 150 years ago, explaining the recent appearance of the stream. Similar work was also performed by \citet{hugh79}.

\indent \citet{hasegawa79}, was the first to propose a potential parent body for the Quadrantids, noticing a similarity between the orbits of the mean Quadrantid stream and comet 1491 I (= C/1490 Y1), recorded in ancient Chinese observations. However, only a parabolic solution was assumed for the orbit of comet 1491 I, due to the low observational accuracy in the position of the comet. Based on the arguable similarity between the orbits of 1491 I and the Quadrantids, the authors concluded that 1491 I had been a periodic Jupiter-family comet, which suffered a very close encounter with Jupiter and was perturbed into a longer period orbit, where the orbital association with its meteoroid stream was lost. 

\indent Assuming that 1491 I was the actual parent of the Quadrantids using the calculated orbital elements of the comet, \citet{Williams93} concluded that the stream was created $\sim$ 5000 years ago. Based on the hypothesis of a very shallow close encounter between comet 1491 I and Jupiter, \citet{Williams93} demonstrated that prior to the encounter with Jupiter, the eccentricity of the orbit of the comet must had been $e \approx 0.77$. The newly derived value for the eccentricity was used for backwards integration of the orbit of the comet to about 5000 years. Then, that epoch was used for the meteoroids ejection whose orbits were integrated forward until the present. The authors argued that the observed mean orbital elements of the stream is consistent with dust particle ejection $\sim$ 5000 years ago. However, the lack of precise orbital elements for 1491 I, along with a hypothesized close encounter with Jupiter, renders the later conclusion uncertain. For a similar work, see also \cite{Lee09}.

\indent Another possible parent of the Quadrantids is the comet 96P/Machholz (formerly P/1986 VIII). \citet{mcintosh90} calculated the secular precession of the orbital elements for the Quadrantids and comet 96P/Machholz and found that the long-term evolution of both orbits is strikingly similar, except for their precession cycles being shifted by a period of 2000 years. The author suggested that the stream was quite old and the phase shift in the precession cycles is due to the differential precession of the orbits of the stream and the comet. Moreover, the author argued that Quadrantids may be a part of a larger complex of meteoroid stream, belonging to comet 96P/Machholz.

\indent \citet{bab92} integrated the orbits of three test particles similar to that of comet 96P/Machholz for 8000 years back in time. Then 20 test particles were ejected from the nucleus of 96P at the epoch of 4500 B.C and integrated forward until 3000 AD. For that period of 7500 years, the authors argued that meteoroids released by 96P can produce eight meteor showers on Earth within one precession cycle of the argument of perihelion $\omega$ of the meteoroids. Six of these showers have been identified as: the Quadrantids, the Ursids, Southern $\delta$ - Aquarids, daytime Arietids, Carinids and $\alpha$ - Cetids. That led the authors to conclude that 96P/Machholz is the parent of the Quadrantid meteoroid stream. For additional and more extensive work, see also \cite{jones93} and \cite{neslusan07}.

\indent \citet{jen97} used $\sim$ 35 doubly-photographed Quadrantids taken in 1995 by the {\it Dutch Meteor Society} ({\it DMS}) \citep{betlem95}, to argue that the age of the central portion of the Quadrantid stream is only $\sim$ 500 years old and the parent may be hidden on a {\it Near Earth Object} ({\it NEO}) - like orbit. With the discovery of 2003 EH$_1$ in 2003 \citep{jen03,mc03}, \citet{jen04} noted the striking similarity between the current orbit of the Quadrantids and the orbit of 2003 EH$_1$ and proposed a sibling relationship.

\indent \citet{wiegert05} estimated an approximate age of 200 years for the core of the Quadrantids, based on the nodal regression rate of the stream and forward integration of meteoroids, released by 2003 EH$_1$ circa 1800 AD. The authors concluded that meteoroids released prior to 1800 AD appear on the sky at much earlier times than the first reported appearance around 1835.

\indent The main goal of this work is to estimate the most probable age of the central portion of the Quadrantid meteoroid stream and its mode of formation (e.g. cometary sublimation vs. asteroidal disruption). We seek to first constrain the approximate formation age assuming 2003EH1 is the parent by first performing backward integrations of high precision Quadrantid meteoroids to compare the orbital similarity between the meteoroids and 2003 EH$_1$. Having established an approximate age from backward integrations we then attempt to simulate the formation of the core of the stream forward in time using the formation epoch found from backward integration and compare with the characteristics of the stream.  However, our intention is not to provide a complete and detailed picture of all physical characteristics of the stream, rather we aim to demonstrate whether the observed overall characteristics of the core of the stream can be explained by assuming a relatively recent (a few hundred years) formation age 
derived from backward integrations of individual meteoroids.

\indent As a test of reliability of our backward integration estimate of the age and formation mode of the stream, we compare the following theoretical and observed characteristics of the stream:

1. The timing of the appearance of the stream on the sky (around 1835 AD).\\
\indent 2. The mean position and spread of the geocentric radiant of the stream.\\
\indent 3. The position of the peak of the activity profile of the core.\\
\indent 4. The width of the activity profile of the core (FWHM $\approx$ 0.6 days).

\indent Throughout this work, we use an approach similar to that of \cite{gust89}.  That author integrated backward in time the orbits of 20 high precision Geminids, along with the parent 3200 Phaethon, and compared the epochs at which the orbits of the Geminids and that of Phaethon intersected. Moreover, he calculated the probable meteoroid ejection speed and location on the orbit of the parent, and concluded that the Geminids are consistent with cometary sublimation that might have taken place on Phaethon around 600 -2000 years ago. For an exhaustive description of the method see also \cite{adol96}.

\section{Asteroid 2003 EH$_1$ and the Quadrantids meteoroid streams}\label{quads&EH1}

Asteroid 196256 (2003 EH$_1$) was discovered in March 2003 by the {\it{Lowell Observatory Near-Earth Object Search}} (LONEOS) and has been designated as a NEO Amor type asteroid. Currently, 2003 EH$_1$ moves on a highly inclined cometary-like orbit with a Tisserand parameter with respect to Jupiter of T$_J$=2.06 but shows no evidence of a cometary activity. 

\indent Recent photometric observations of the asteroid 2003 EH$_1$ give an absolute visual magnitude $H = 16.2$ mag, given by the NASA's JPL Small Body Database Browser (\url{http://ssd.jpl.nasa.gov/sbdb.cgi}). The diameter can be found using the expression given by \citet{chesley02}:

\begin{equation}\label{eq:diameter}
 D(km) = 1329 A^{-0.5}\times 10^{-0.2H}, 
\end{equation}

where $A$ is the geometric albedo and $H$ is the absolute visual magnitude of the asteroid, respectively. A typical albedo value for asteroids (depending on the spectral class of the asteroid) is 0.04 $\textless$ $A$ $\textless$ 0.4 \citep{haris89}, which using Eq.~\ref{eq:diameter} yields a diameter of 1.2 km $\textless$ $D$ $\textless$ 3.8 km.

\begin{figure}[h!]
\begin{center}
\advance\leftskip -3cm
\advance\rightskip -3cm
\includegraphics[width=1.0\linewidth, angle=-90]{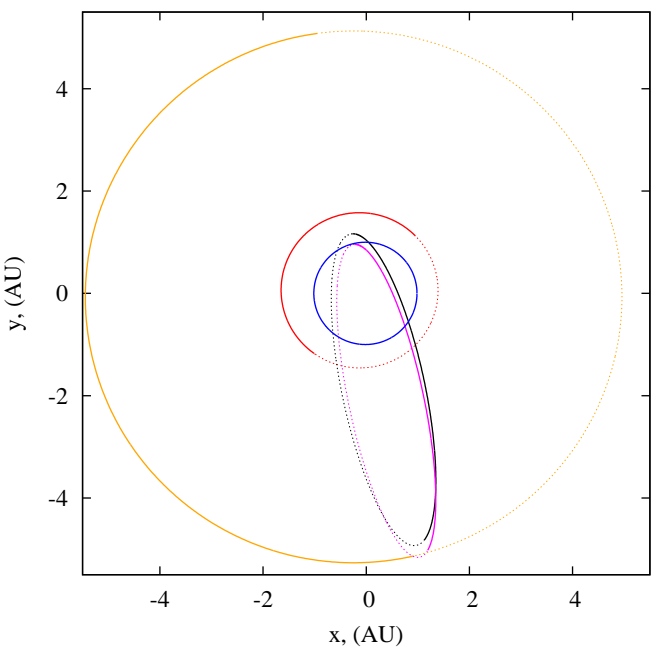}
\caption{\footnotesize{Orbits of asteroid 2003 EH$_1$ (black line) and the mean Quadrantid stream \citet{jen97} (magenta line), viewed from above the ecliptic plane. The orbit of Earth is indicated with blue, Mars with red and Jupiter with orange colors, respectively. The portions of the orbits, below the ecliptic are denoted with a dashed line. The ascending nodes of the orbits of both, the mean Quadrantids and 2003 EH$_1$ are located close to the Jupiter's orbit.}}
\label{2003EH1}
\end{center}
\end{figure}

\indent The present orbit of the asteroid 2003 EH$_1$ is presented in Fig.~\ref{2003EH1}, along with the orbits of the Earth, Mars and Jupiter and the mean Quadrantid orbit. The orbital elements of 2003 EH$_1$ and the mean orbit of the Quadrantids are given in Table~\ref{tab1}, along with other previously suggested parents, C/1490 Y1 and 96P/Machholz. The orbital elements of the mean Quadrantid stream are taken from \citet{jen97}.

\begin{figure}[h!]
\begin{center}
\advance\leftskip -3cm
\advance\rightskip -3cm
\includegraphics[width=0.7\linewidth,angle=-90]{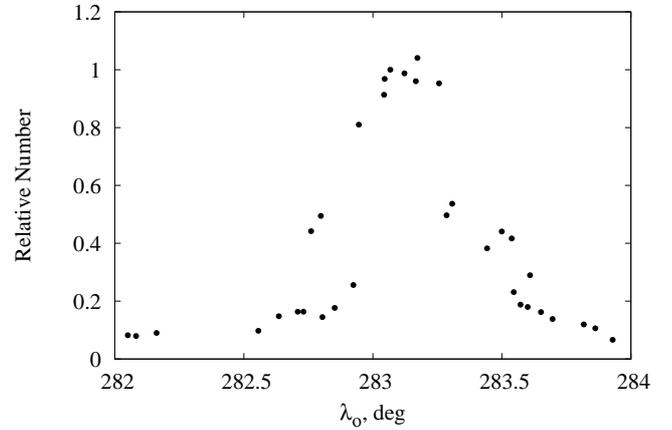}
\caption{\footnotesize{The average visual activity profile of the Quadrantid meteor shower. The profile is a stack of a few favorable observations between 1986 - 1992 \citep{Ren93}. The peak of the activity is centered around $\lambda_{\odot}$ = 283.2$^{\circ}$ for equinox of J2000.0.}}
\label{visual-obsprof}
\end{center}
\end{figure}

\indent The observed visual activity profile of the Quadrantid meteor shower is presented in Fig.~\ref{visual-obsprof}. It is a composite (average) of activity profiles, as deduced from visual observations of Quadrantids, in the years of 1986, 1987, 1989, 1990 and 1992 \citep{Ren93}. The core of the activity profile is easily seen to be less than a day wide, with a FWHM $\approx$ 0.5 days.

\begin{figure}[h!]
\begin{center}
\advance\leftskip -3cm
\advance\rightskip -3cm
\includegraphics[width=0.7\linewidth,angle=-90]{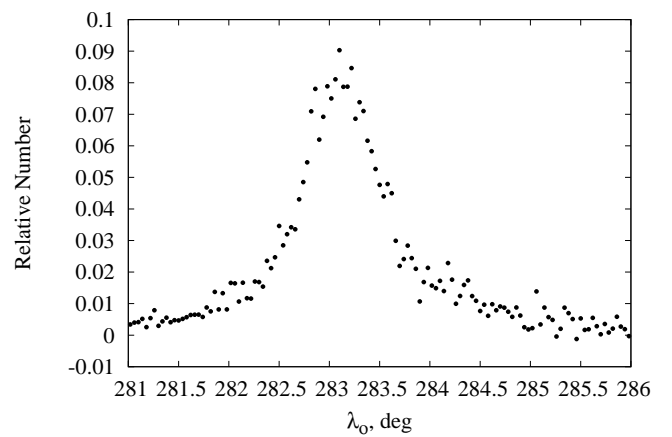}
\caption{\footnotesize{The average radar activity profile of the Quadrantid meteor shower as deduced by the {\it{Canadian Meteor Orbit Radar}} ``CMOR``. The profile is an average profile of all good quality observations between 2002 - 2014.}}
\label{radar-CMOR}
\end{center}
\end{figure}

\indent Fig.~\ref{radar-CMOR} shows the average radar activity profile of the Quadrantids (milligram size particles), as observed by the {\it{Canadian Meteor Orbit Radar}} ``CMOR'' (see Section~\ref{obs} for details). The peak of the activity, located at $\lambda_{\odot}$ = 283.1$^{\circ}$ for equinox of J2000.0, occurs slightly before the visual meteors peak, but is in agreement within uncertainty of the two (averaged) profiles. The FWHM corresponds to approximately 0.8 days, slightly wider than the visual activity profile (see Fig.~\ref{visual-obsprof}).

\indent As shown in Fig.~\ref{2003EH1}, the stream of particles producing the Quadrantids intersect the orbit of the Earth at 
their descending node, whereas the ascending node is located near the orbit of Jupiter. We emphasize that the mean Quadrantid 
orbit, as observed from the Earth, does not necessarily represent the orbit of the entire physical stream, but rather only that 
part that intersects the Earth. Meteoroid streams are usually dispersed and not all members intersect the Earth, so it is not 
possible to know the orbits of that portion of the stream that does not physically intersect the Earth's path. Therefore we refer 
to the mean Quadrantid stream as the mean of that portion that interacts with the Earth's atmosphere.

\begin{table*}[!htpb]
\centering
{\footnotesize
\begin{tabularx}{\textwidth}{ p{3.5cm} p{1.7cm} p{1.2cm} p{1.2cm} p{1.2cm} p{1.2cm} p{1.2cm} p{1.2cm} p{1.2cm}}
 \toprule[1pt]
 Object & Epoch (UT) & a (AU) & e & q (AU) & Q (AU) & i (deg) & $\omega$ (deg) & $\Omega$ (deg)\\\hline\\
 2003 EH$_1$               & 04 Nov. 2013 & 3.122 & 0.619 & 1.189 & 5.054 & 70.876 & 171.354 & 282.962\\
 96P/Machholz           & 06 Sep. 2013 & 3.034 & 0.959 & 0.124 & 5.944 & 58.312 & 14.757  & 94.323\\
 1490 Y1                & 26 Dec. 1490 & ---   & 1     & 0.737 & ---   & 51.65  & 129.84  & 295.89\\
 Quadrantids & ---          & 3.14  & 0.688  & 0.98 & 5.3   & 71.5   & 171.2   & 283.3\\
\bottomrule
 
\end{tabularx}
}
\hfill{}
\caption{\footnotesize{The osculating orbital elements of asteroid 2003 EH$_1$, comet 96P/Machholz, comet 1490 Y1 and the mean 
Quadrantid orbit (J2000.0), as given in the NASA's JPL Horizons System (\protect\url{http://ssd.jpl.nasa.gov/sbdb.cgi}). The 
orbital elements of the mean stream are taken from \citep{jen97}}. $Q$ and $q$ in the table stand for the aphelion and perihelion 
distances respectively.}
\label{tab1}
\end{table*}

\indent It is also evident that present orbit of the Quadrantid meteoroid stream and that of asteroid 2003 EH$_1$ are strikingly similar. However, the proximity of the parent and stream orbits is not a sufficient, nor necessary condition for a relation between the two. For example, initially close orbits (e.g. stream and parent) may follow dramatically different dynamical evolution due to the differential gravitational perturbations and solar radiation forces. Conversely, two completely different orbits may end up into similar ones due to the aforementioned effect and result in a "false" sibling relationship between the two orbits. Thus, robust models of dynamical evolution of streams must be employed. 

\indent \citet{jen04} were the first to notice the similarity between 2003 EH$_1$ and the Quadrantids and proposed the former as the parent of the stream. However, the lack of cometary activity on 2003 EH$_1$ and the nature of its orbit make it a strong candidate for a recently dormant or defunct comet. The real nature, whether cometary or asteroidal, is unknown due to the lack of favorable returns and observations, rendering the formation mechanism of the Quadrantids uncertain.  

\indent The nature of 2003~EH$_1$ will determine the underlying mechanism by which it produced the Quadrantids. While cometary
activity is driven by the sublimation of water ice, asteroids may also shed material by other processes. For example, 
\citet{jewitt12} argued that the mass loss mechanism from the surface of 3200 Phaethon, another seemingly dormant or extinct comet, producing the Geminids meteor shower, is due to solar radiation pressure sweeping.  

\indent The inactivity of 2003 EH$_1$ points perhaps to an asteroidal nature. However, recent spectroscopic and double-station observations of 51 Quadrantids, revealed that unlike the Geminids the Quadrantids are not as depleted in volatiles \citep{koten06}. In fact, the authors argued that the Quadrantids fall in a category between meteoroids of cometary and asteroid origin, and concluded that the parent 2003 EH$_1$ must be a dormant comet. Given 2003~EH$_1$'s comet-like orbit, we will here assume that material is released from it by traditional cometary activity in our forward modeling.

\section{Meteoroid ejection model}\label{ejecmodel}

Besides meteoroid ejection due to cometary sublimation, there are a few other potential mechanism of mass shedding from the surface of small bodies (e.g. \citep{jewitt12}) such as impact ejection, rotational instability, electrostatic forces, thermal fracture and radiation pressure sweeping. However, in this work we limit ourselves to cometary sublimation as the most likely mechanisms of formation of the narrow core of the Quadrantids.

\subsection{Cometary sublimation}

Presently, the meteoroid production mechanism due to cometary volatiles sublimation is relatively well understood, see e.g. \citep{whip51,jones95,crifo97,Mawil,Hugh00}. All meteoroid ejection models to a great extent share the same physical concepts although with slight modifications. 

\indent Perhaps the very first meteoroid ejection model came with the pioneering work done by Fred Whipple \citep{whip51}, when he proposed his "icy-conglomerate" (''dirty snowball'') comet nucleus model. In his model, the comet nucleus consists of frozen volatiles - ices with solid refractory particles embedded within the ice. Once the comet nucleus is close enough to the Sun, e.g 3 AU \citep{Delsemme82}, the cometary ices begin to sublimate, releasing and dragging along the embedded dust particles. In Whipple's model, the meteoroids leave the surface of the comet with speeds:

\begin{equation}\label{eq:ej-whipple}
 V_{\mathrm{ej}}=25.4\: r^{-1.125}\rho^{-1/3}R_c^{1/2}m^{-1/6} \; \text{(m/s)}
\end{equation}

where the $r$ is the heliocentric distance in AU, $\rho$ is the density of the meteoroid in kg/m$^3$, $m$ is the mass of the meteoroid in kg and $R_c$ is the radius of the comet in km. However, one major disadvantage of Whipple's model, later improved by \citet{jones95}, is the assumption of blackbody limited temperature of the comet nucleus. Moreover, \citet{jones95} considered an adiabatic expansion of the escaping gas, i.e the sublimation of the ices results in cooling of the comet's surface. The terminal speed of the particles, leaving the surface of the comet, according to \citet{jones95} is:

\begin{equation}\label{eq:ej-jones}
 V_{ej}=29.1\: r^{-1.038}\rho^{-1/3}R_c^{1/2}m^{-1/6} \; \text{(m/s)}
\end{equation}

where the variables and the units are the same as in Eq.\ref{eq:ej-whipple}.

\indent One major shortcoming of Jones' model is the assumption that the entire nucleus is active. Direct imaging of the nucleus of comet 1P/Halley, by {\it Giotto} and {\it Vega} comet probes, showed that the sublimation of the nucleus is confined to a small fraction of the cometary surface. Nonetheless, a close inspection of Eq.~\ref{eq:ej-jones} shows that the latter expression is very similar to that derived by Whipple. This indicates that accounting for the adiabatic expansion of the escaping gases and eliminating the blackbody nucleus limitation yields little difference in the final result for the terminal velocity of the ejected meteoroids. 

\indent More recently, in their work of modeling the dynamical evolution of the Perseid meteoroid stream, \citet{BJ98} used a
slightly modified version of the \citet{jones95} ejection speed as a function of the heliocentric distance. Moreover, the 
probability of particles having ejection speed $P(V-V_{ej})$ was assumed to be a parabolic distribution, where the meteoroid 
ejection directions are distributed isotropically on the sunlit hemisphere. Regarding the meteoroid production, we follow 
\citet{kresak76}, where the production rate is uniform in true anomaly ($\nu$). According to \citet{BJ98}, the terminal speed at 
which the meteoroids leave the surface of the comet is given by:

\begin{equation}\label{eq:ej-bj}
 V_{ej}=10.2\:r^{-0.5}\rho^{-1/3}R_c^{1/2}m^{-1/6} \; \text{(m/s)}
\end{equation}

with ejection probability distribution:

\begin{equation}\label{eq:ejdist-bj}
 P(V-V_{ej})=1 - \left(\frac{V}{V_{ej}}-1\right)^2 \; \text{with}\; 0<V<2V_{ej}\; \text{and 0 outside} 
\end{equation}

where $P(V-V_{ej})$ is the probability of finding a meteoroid with ejection speed $V$.

\indent In the beginning of 21$^{st}$ century, several interesting works appeared in which authors attempted to infer the ejection speeds of meteoroids causing a meteor outburst at the Earth. Observations of the profile of the shower, given the meteoroids were released at a certain perihelion passage of the parent, can be used to analytically solve for the ejection conditions from the parent. The method has extensively been applied to the 1999 Leonid outburst (see e.g. \citep{gockel00,mawill01,muller01}). This approach, however, has its limitations i.e the meteoroids producing the outburst must be large (mm - cm) and originate from relatively young meteoroid trails (a few orbital periods) \citep{asher08}. The ejection conditions for much older meteoroids is more difficult (almost impossible) to be deduced as planetary perturbations, over several orbital periods, become significant thus rendering the recovery of ejection conditions impossible. This method, however, is not suitable for our simulations for a few reasons: first the most probable age of the Quadrantids, based on our backward integrations, (200-300 years $\approx$ 35-55 orbital revolutions of 2003 EH$_1$) renders it impossible to apply the above method. Secondly, we do not have a priori knowledge on the physical properties of the nucleus, as the cometary activity may vary significantly from comet to comet. This method, however, is worth for future investigation. 

\indent Throughout this work, for our forward modeling, we choose to use the above (Eq~\ref{eq:ej-bj} and Eq~\ref{eq:ejdist-bj}) 
ejection model by \citet{BJ98}. We assume that 2003 EH$_1$ has a mean density of $\rho_{c} =$ 800 kg/m$^3$ - a typical value for 
 a comet nucleus \citep{weissman04}, while for the radius of the parent body we use an average value of $R_c =$ 1 km, see 
Section~\ref{quads&EH1}. The meteoroids, on the other hand, are modeled as spherical grains of density $\rho=$ 1900 kg/m$^3$ 
\citep{bab02}, where only the radius of the particles is allowed to vary (Section~\ref{forward}). Due to the present lack of 
cometary activity on 2003 EH$_1$, we assume it to be relatively depleted of volatiles such as: $CO$, $CO_2$, $CH_4$, $NH_3$ etc. 
and thus comprising mostly of water ice (if any) and embedded refractory material (meteoroids). The water ice begins to sublimate 
at a heliocentric distance of roughly $r=$~3 AU \citep{Delsemme82}, thus the meteoroids, in 
our simulations, will be released from 2003 EH$_1$ on an arc of the orbit within 3 AU from the Sun, centered at the perihelion of 
2003 EH$_1$. For example, assuming an orbit similar to that of 2003 EH$_1$, see Tab~\ref{tab1}, and meteoroid 
ejection model that of \citet{BJ98}, the magnitude of the ejection speed of a meteoroid, with bulk density of $\rho=1900$ 
kg.m$^{-3}$ and mass $m=7.6\times10^{-9}$kg (corresponding to a radius of 1 mm), will be $V_{ej}\sim~18$~m/s at $r$= 1 AU, 
$V_{ej}\sim~8.8$~m/s at $r$= 2 AU and $V_{ej}\sim~5.8$~m/s at $r$= 3 AU.

%

\section{Observational Data}\label{obs}

In our study, we use the orbital data of eight high-precision Quadrantids, photographed from multiple stations by the Czech part of the {\it European Fireball Network} (``EN''), \citep{spurny94, spurny07}. In addition to these bright photographic Quadrantids, we also use five core radar Quadrantids observed by the {\it Canadian Meteor Orbit Radar} (CMOR) \citep{jones05}, which sample the smaller  members of the stream (milligram size). These Quadrantids were backward integrated and their orbit compared with that of 2003 EH$_1$ to search for epochs of close encounters which might indicate when the material forming the young core of the stream was released.

\subsection{Photographic Quadrantids}

Five out of the eight photographic Quadrantids pertain directly to the narrow portion of the stream, as they have been observed within a degree from the peak ($\lambda_{\odot}=283.2^{\circ}$), which we refer to as ``core Quadrantids''. The remaining three lie slightly outside of the core of the stream with two extreme cases being Dec. 31 ($\approx 3^{\circ}$ from the peak) and Jan. 8 ($\approx 4.2^{\circ}$) (see Table~\ref{tab2}). Meteors observed a few degrees away from the peak of the Quadrantid shower are referred to as ``non-core Quadrantids''.  

\indent These bright meteors are of an exceptional importance in our simulations, as their photometric mass (gram-kilogram) implies that the effect of the solar radiation pressure on these meteors, given the timescale of our simulations of a few hundred years, would be negligible. However, the mass of each individual photographic meteoroid can be uncertain by a factor of 3 due to our poor knowledge of the value of the {\it luminous efficiency} $\tau$, used for the estimation of the ablated mass. Thus, in order to ensure that the effect of the solar radiation pressure has properly been taken into, we consider a wide range in the $\beta$ - parameter, where $\beta = F_R/F_G$ is the ratio of the solar radiation pressure to the solar gravity (Section~\ref{ns}). The assumed values for the $\beta$ - parameter are: $0, 10^{-3}, 10^{-4}, 10^{-5}$. The aim is to encompass the entire possible range in $\beta$, because of the a priori uncertainty in the mass of the meteoroids. The magnitude of $\beta$ can be expressed 
as:

\begin{equation}\label{eq:burns1}
  \begin{aligned}
 & \beta = \frac{F_R}{F_G} \approx 3.54\times 10^{-5}\frac{Q_{pr}}{m_{o}^{1/3}\rho^{2/3}}, \quad \text{(``cgs`` units)}\\ 
 & \beta = \frac{F_R}{F_G} \approx 3.54\times 10^{-4}\frac{Q_{pr}}{m_{o}^{1/3}\rho^{2/3}}, \quad \text{(``SI`` units)}
   \end{aligned}
\end{equation}

where, for instance, a spherical meteoroid of mass 1 gram and bulk density of 
$\rho=1.9$ g/cm$^{3}$ has $\beta$ $\approx 2.3 \times 10^{-5}$. It is evident that the larger meteoroids are less 
susceptible to the solar radiation pressure and in fact, for bolides (kilogram size) $\beta$ can safely be assumed to be zero.

\indent Except the photometric mass, the quality of the data for each individual photographic Quadrantid is exceptionally good, with the uncertainty in the speed (before the atmospheric entry) ranging between $\approx$ 0.05-0.5 \%, whereas the radiant position is known to $\approx$ 0.02$^{\circ}$. Unlike other authors who use the mean orbital elements of the Quadrantids as the starting point for integrations, we use these high precision observable and measurable quantities i.e the geographic coordinates and speed of a point on the meteor trajectory and the radiant coordinates. These quantities can directly be translated into geocentric state vectors for each observed Quadrantid providing a high precision initial osculating orbital elements used in our backward simulations. 

\begin{table*}[!htpb]
 \centering
{\footnotesize
\begin{tabularx}{\textwidth}{p{1.8cm}p{1.5cm}p{1.1cm}p{1.1cm}p{0.8cm}XXXXXXXXXX}
\toprule[1pt]
Meteor Number & Date& $\lambda$ & $\phi$ & {\it{h}} & \it{m$_{\circ}$} &R.A & Dec. &  {\it{V$_{\infty}$}} & $a$& $e$& $i$& $\omega$& $\Omega$ \\ 
 & (UT)& ($^{\circ}$) & ($^{\circ}$) & (km) & (g)& ($^{\circ}$) & ($^{\circ}$) & (km/s)& (AU) & & ($^{\circ}$)&($^{\circ}$)&($^{\circ}$)\\
 \cline{7-14}\\
 & & & & & &\multicolumn{8}{c}{(J2000.0)}\\
\\\hline
\textbf{Core} & & & & & & & & & & & &\\
\textbf{Quadrantids} & & & & & & & & & & & &\\
\cline{1-1}\\

EN030109& 2009-Jan-3 & 11.1406    & 48.95368     & 93.29     & 1200 & 227.38 & 52.03 & 42.57 & 2.86 & 0.656& 72.18& 176.66 &283.3930 \\
       & 18:19:14.461 & $\pm$0.0003& $\pm$0.00005 & $\pm$0.02 &--& $\pm$0.04 & $\pm$0.01& $\pm$0.02 & $\pm$0.01 &$\pm$0.002& $\pm$0.03& $\pm$0.06& --\\\\
       
EN040180 & 1980-Jan-4 & 16.05305     & 50.12508    & 93.30       & 1.9 & 232.28   & 50.40 & 41.92 & 2.79& 0.650& 70.98& 168.51& 283.7938\\
         & 17:07:34    & $\pm$0.00004  &$\pm$0.00001& $\pm$0.01   & --  & $\pm$0.02 &$\pm$0.01& $\pm$0.10 & $\pm$0.05& $\pm$0.005& $\pm$0.10& $\pm$0.08& --\\\\

EN030189 & 1989-Jan-3 & 14.9824     & 48.6881    & 93.53       &  2.3 & 235.90  & 49.67     & 40.3     & 2.82 & 0.656 & 66.6& 164.6& 283.7938\\
         & 4:27:10    & $\pm$0.0001 &$\pm$0.0001 & $\pm$0.01   & --& $\pm$0.07  & $\pm$0.03 & $\pm$0.3 & $\pm$0.15& $\pm$0.02& $\pm$0.3& $\pm$0.3& --\\\\
         
EN040192 & 1992-Jan-4 &  15.98771 & 49.92328 & 91.33        & 80  & 227.38    & 49.96     & 42.54 & 2.71& 0.637& 71.87& 174.45 & 283.1019\\
         & 2:48:30 & $\pm$0.00002  &$\pm$0.00001 &$\pm$0.01 & --  & $\pm$0.02 & $\pm$0.01 & $\pm$0.07 & $\pm$0.03& $\pm$0.004& $\pm$0.08& $\pm$0.04 & --\\\\

EN040111 & 2011-Jan-4 &  15.98771     & 49.92328    & 91.33       & 70 & 232.53  & 49.67   & 41.1 & 2.21      & 0.642      & 68.7& 167.7 & 283.2498\\
         & 3:19:43 &$\pm$0.00002  &$\pm$0.00001 &$\pm$0.01    & --& $\pm$0.03  &$\pm$0.01 & $\pm$0.3 & $\pm$0.14& $\pm$0.018& $\pm$0.3& $\pm$0.2 & --\\\\\\

\textbf{Non-core} & & & & & & & & & & & &\\
\textbf{Quadrantids} & & & & & & & & & & & &\\
\cline{1-1}\\
EN311210 & 2011-Dec-31 &  14.39278     & 47.37932    & 93.71       & 650 & 226.77  & 51.11   & 42.32   & 2.88 & 0.659& 70.0& 173.10 & 279.1733\\
         & 3:21:49.71& $\pm$0.00040  &$\pm$0.00023 & $\pm$0.03   & --& $\pm$0.14&$\pm$0.02& $\pm$0.17  & $\pm$0.09& $\pm$0.011& $\pm$0.2& $\pm$0.18 & --\\\\

EN060111 & 2011-Jan-6 &  16.51110     & 48.29288    & 80.18       &  30 & 227.02  & 49.86    & 42.6 & 2.47& 0.602& 71.9& 178.72 & 285.3437\\
         &4:36:52.65 & $\pm$0.00035  &$\pm$0.00025 &$\pm$0.05    & --& $\pm$0.08  &$\pm$0.08 & $\pm$0.3 & $\pm$0.12& $\pm$0.019& $\pm$0.3& $\pm$0.18 & --\\\\

EN080179 & 1979-Jan-8 &  15.61535     & 48.96701    & 98.02       &  60 & 233.53  & 48.55   & 43.21   & 3.73& 0.738& 71.61& 170.41 & 287.5665\\
         &3:43:55 &  $\pm$0.00007  &$\pm$0.00005 &$\pm$0.02    & --& $\pm$0.03 &$\pm$0.01 & $\pm$0.18 & $\pm$0.17& $\pm$0.012& $\pm$0.19& $\pm$0.10 & --\\
\bottomrule
\end{tabularx}
}
\hfill{}
\caption{\footnotesize{Data for eight high precision Quadrantids, photographed from multiple stations by the Czech part of the {\it European Meteor Network}. The table is divided into two parts, {\it{core Quadrantids}} - observed within a day of the peak of the shower and {\it{non-core Quadrantids}} - observed within a few days of the peak. 
\indent The first column is the meteor identification number. The second column is the date and time of observation in (UT). The third, fourth and fifth columns are the the geographic position of a point on the meteor trajectory (geographic longitude $\lambda$, geographic latitude $\phi$ and altitude {\it h} with respect to the sea level). The mass of the meteoroid is given in the sixth column. The seventh and eighth columns are the geocentric radiant position for (J2000.0) - Right Ascension $\alpha$ and Declination $\delta$, followed up by the pre-atmospheric speed $V_{\infty}$ of the meteoroid (ninth column). The last five columns correspond to the heliocentric orbital elements (J2000.0) - semi-major axis ($a$), eccentricity ($e$), inclination ($i$)
, argument of perihelion ($\omega$), and longitude of ascending node ($\Omega$). All uncertainties are one $\sigma$.}}

\label{tab2}
\end{table*}

\subsection{Radar Quadrantids}

\indent Radar observations of the Quadrantid meteor shower have been reported by several authors since 1947 \citep{almond52}; with some major studies carried out by e.g. \cite{almond52, mill53, poole72}. Radar observations can be conducted even during daytime and overcast weather, and can detect meteoroids of much smaller sizes (milligram size) than those responsible for the visual meteors (gram-kilogram). Radar observations of the Quadrantids complement visual measurements as the shower peaks in early January when weather conditions are often poor in the northern hemisphere.

\indent CMOR is a triple-frequency multi-station meteor radar system (17.45, 29.85, 38.15 MHz), with the main station located near Tavistock, Ontario, Canada (43.264 N, 80.772 W), and five other remote sites, recording $\sim$ 4000 orbits per day. Using the time delay for common echoes between stations and echo directions of the received signal at the stations, a meteor's speed and atmospheric trajectory can be determined. While single station operation registers echoes (where only echo direction and ranges from the main site are measured) and provides some physical information about the meteoroid, a multi-station detection allows for the determination of a meteoroid's heliocentric orbit (for details, see e.g \citet{webster04, jones05, weryk13}).

\indent During the 2013 campaign, CMOR measured $\approx$ 1440 radar Quadrantids orbits (with masses of the order $10^{-3}$ -  $10^{-4}$ grams). Out of that list, we have selected 31 Quadrantids with heights above 100 km, to minimize atmospheric deceleration, and fractional error in the semi-major axis $da/a$ $\textless$ 0.1. Of the 31 preliminary selected Quadrantids, we extracted only 5 with highest signal-to-noise ratio echoes, as well as the least uncertainty in the atmospheric velocity $\approx$ 1\% where all uncertainties were found using a Monte Carlo routine approach (\citep{weryk12}). Table~\ref{tab3} lists the observed geocentric quantities and radiants for these high-quality radar orbits.

\begin{table*}[htpb]
 \centering
{\footnotesize
\begin{tabular*}{\textwidth}{lccccccccccccc}
\toprule[1pt]
Date & Time &  $\lambda$ & $\phi$    & {\it{h}} &  \it{m$_{\circ}$} & $\lambda_{\odot}$ & $\alpha$  & $\delta$  & {\it{V$_{\infty}$}} & $a$ & $e$ & $i$ & $\omega$ \\
     & (UT) &  ($^{\circ}$) & ($^{\circ}$) & (km)     &  $\times$ 10$^{-3}$ (g) & ($^{\circ}$) & ($^{\circ}$) & ($^{\circ}$) & (km/s)    & (AU) & ($^{\circ}$) & ($^{\circ}$)& ($^{\circ}$) \\
\cline{7-14} \\     
& & & & & & \multicolumn{8}{c}{(J2000.0)}\\\hline
  
20130103 & 09:11:22  & -82.23    & 43.22      &   100.863   &  1.1      & 282.98     &  224.2     & 46.9          &  42.2       & 1.81      & 0.466        & 74.44      & 173.4\\

         &           &           &            &             &             &            &            &             & $\pm$ 0.7   &$\pm$0.07  & $\pm$0.023   & $\pm$0.41  & $\pm$1.7\\\\
         
20130103 & 20:08:57  &  -79.47   & 43.55      & 101.482    & 1.7        & 283.44     &230.5       & 48.4          & 40.2        & 2.01      & 0.512        & 69.84      & 168.1\\
	 &           &           &            &            &            &            &            &               & $\pm$ 0.8   &$\pm$0.04  & $\pm$0.011   & $\pm$0.15  & $\pm$0.5\\\\
	 
20130103 & 20:26:56  &  -80.47   & 42.83      & 100.678    &  0.6       & 283.45     & 230.1      & 48.7          & 40.9        & 2.22      & 0.559        & 70.56      & 169.8\\
	 &           &           &            &            &            &            &            &               & $\pm$ 0.1   & $\pm$0.06 & $\pm$0.012   & $\pm$0.2   & $\pm$0.3\\\\
	 
20130103 & 20:50:27  &  -78.52   & 44.25      & 103.118    &  5.5       & 283.47     & 224.1      & 46.8          & 41.5        & 1.87      & 0.476        & 73.43      & 171.6\\
	 &           &           &            &            &            &            &   	  & 		  &$\pm$ 0.1    & $\pm$0.11 & $\pm$0.031   & $\pm$0.59  & $\pm$2.1\\\\
	 
20130104 & 06:16:11  &  -80.95   & 42.91      & 103.046    &  5.7 	& 283.87     & 235.1      & 53.6          & 39.7        & 1.91      & 0.488        & 68.23      & 171.2 \\
	 &           &           &            &            &            &            &  	  & 		  & $\pm$ 0.3 	& $\pm$0.07 & $\pm$0.021   & $\pm$0.37  & $\pm$0.4\\

\bottomrule
\end{tabular*}
}
\hfill{}
\caption{\footnotesize{Data for five high quality radar Quadrantids, belonging to the narrow core of the stream,observed by CMOR during 2013. The columns represent the date of observation, time, geographic longitude $\lambda$, latitude $\phi$ and height $h$ above the ground of point on the meteor trajectory, mass m$_{\circ}$ of the meteoroid, the solar longitude at the time of observation $\lambda_{\odot}$, geocentric right ascension $\alpha$ and geocentric declination $\delta$ of the radiant position and the speed V$_{\infty}$ of the meteoroid above the Earth's atmosphere. The last four columns are the computed heliocentric orbital elements of the meteoroid, semi-major axis $a$, eccentricity $e$, inclination $i$ and argument of perihelion $\omega$. The average uncertainty in the radiant position is $\pm$ 0.2$^{\circ}$}.}
\label{tab3}
\end{table*}

\indent While the velocity is known to $\approx$ 1\% in our data set, the mass estimate of the meteoroid is more uncertain. The ablated mass is a function of the speed of the meteoroid $V$ in the atmosphere and the {\it electron line density} $q$ \citep{verniani73}, created by collisions of the evaporated meteoric atoms with atmospheric molecules. The major uncertainty in the mass arises from the uncertainty in the electron line density $q$, which translates into an error in the mass of a factor of 3 \citep{weryk12}. For a bulk density of a meteor of $\rho = $1.9 g/cm$^3$ \citep{bab02} and mean value for the mass of radar Quadrantids $m \approx 3\times10^{-3}$ grams, see Table~\ref{tab3}, this yields for a typical $\beta$ - parameter for a radar meteor $\beta \approx 1.6\times 10^{-4}$, if we use for the {\it scattering efficiency} $Q_{pr} = 1$. Even if the mass for a radar Quadrantid was uncertain by a factor of 3, that would translate in error for the  $\beta$ - parameter $\beta \approx 2 \times 10^{-4}$ 
and  $\beta \approx \times 10^{-4}$ for an under and overestimated mass, respectively. These last results confirm that we can safely use values for $10^{-5} \textless$ $\beta$ $\textless 10^{-3}$, in order to encompass the entire possible mass range of the radar Quadrantids.

\section{Numerical Simulations}\label{ns}

\indent We explicitly assume that the parent body of the Quadrantids is the asteroid 2003 EH$_1$. For each individual Quadrantid, (see Table~\ref{tab2} and~\ref{tab3}), we create 10$^4$ hypothetical clones with orbits similar to each individual Quadrantid (see Section~\ref{clones}). We then numerically integrate the equations of motion of all clones, along with the parent, backwards in time (Phase 1). Once the most probable formation epoch has been established, via these back integrations, then that time window is used for meteoroid ejection and forward in time integration of test particle orbits until the present (Phase 2).

\subsection{The ``clones``}\label{clones}

\indent The key point is to integrate the equations of motion of a large number of hypothetical meteoroids (clones) of a Quadrantid observed in the present epoch, along with the presumed parent body, backwards in time and to statistically attempt to determine the epoch of minimum distance between their orbits. This approach is more reliable rather than trying to locate the nearest epoch, back in time, when the parent and the meteoroid actually physically intersect. The latter event is very unlikely to occur, due to the inherent uncertainties in the position and velocity of the bodies, as well as computational errors (truncation and round-off) throughout numerical integrations. Even though the two bodies are seen to intersect in the simulations, that does not necessarily imply that we have identified the epoch of meteoroid release, due to the reasons mentioned above. A simple example can be used to demonstrate that: Even one has calculated the semi-major axis $a$ of a meteoroid orbit with an uncertainty of, e.g $a \pm da = 2.855 \pm 0.013$ AU (as in the case of meteor EN030109 Table~\ref{tab2}), over several revolutions of the 
meteoroid about the Sun, the uncertainty in its position will dramatically increase. Using and differentiating Kepler's third law, it can be demonstrated that; 

\begin{equation}\label{eq:beta}
 dT = \frac{3a^{1/2}da}{2\sqrt{1-\beta}},
\end{equation}
  
where $dT$ is the absolute uncertainty of the period of an orbit around the Sun. This yields a relative uncertainty of $dT/T \approx 7 \times 10^{-3}$ even for $\beta=0$, corresponding to $dT \approx 12$ days per period or one orbit within $\approx 700$ years, comparable to the time scales considered here. It is noteworthy that absolute uncertainty in $dT$ increases for finite values of $\beta$. Thus, the data is not precise enough to determine when the meteoroid and the parent actually intersect: such studies will have to await even higher quality measurements. Nonetheless, we can use the intersection of the meteoroid and parent orbits to examine when such ejection could have occurred, as this is a necessary condition for meteoroid stream production.

\indent For our study, for each observed Quadrantid we create 10$^4$ hypothetical clones, randomly selected from a six-dimensional Gaussian distribution of the six directly observed and measured quantities: geocentric radiant coordinates ($\alpha$, $\delta$), the geographic longitude, latitude and altitude ($\lambda, \phi$ and $h$) and the observed velocity at a point on the meteor trajectory ($V_{obs}$). These values can be translated directly into the initial heliocentric state vectors \{$V_{ix}, V_{iy}, V_{iz}$\} and \{$X_i, Y_i, Z_i$\} of each clone and are used as initial conditions for our backwards numerical integrations of the equations of motion. Each dimension of the six-dimensional Gaussian, is centered around the nominal values with one standard deviation 1 $\sigma$ being equal to the magnitudes of the uncertainties of the initial variables (see Table~\ref{tab2}). However, at the time of our simulations, we do not have information about the covariance between the errors in our initial condition 
quantities, thus we assume that the errors are uncorrelated, which may not be the case.

\subsection{Phase 1: Backward integrations}\label{backward}

\indent The equations of motion of each clone along with the parent are then integrated backwards in time for 1000 years. The length of the backward integrations of 1000 years is chosen on the basis of the assumption of relatively young age of the central portion of the Quadrantid meteoroid stream \citep{jen97, wiegert05} and because we found that similar Quadrantid orbits in the current epoch integrated back in time, do not start to significantly disperse until $\approx$ 1300 AD.
In addition, we aimed to check whether the age of the core of the Quadrantids could be as old as 500 - 1000 years as previously argued by some authors e.g. \citep{jen97, Williams93}. We note that integration for 1000 years back in time is beyond the Lyapunov time, however the mean orbital elements of the clones of 2003 EH$_1$, at the epochs of 1000 AD and 1500 AD, are used for forward integrations in order to compare the theoretical characteristics of core of the Quadrantid stream against the observed ones. Effectively, we explore whether the observed characteristics of the core of the Quadrantids can be explained by meteoroid ejections from 2003 EH$_1$ circa 1000 AD or 1500 AD. However, the results from forward integrations should be treated with caution, in particular when the initial orbital elements of the parent have been selected from backward integration beyond the Lyapunov time. Therefore, the results must be treated from a statistical point of view. 

\indent Throughout the integrations, we consider the perturbations on the meteoroids and 2003 EH$_1$ from each planet and also 
account for the planets' mutual interaction. We used the JPL's ''DE405'' version of the planetary ephemeris for generating the 
initial positions of planets. The mutual interactions among the meteoroids and 2003 EH$_1$ are neglected, i.e they are considered 
as test particles. Throughout the integrations we use Everhart's RADAU algorithm \citep{ever85}. During the first 30 days of the 
integration of the orbits of the clones, i.e. when the meteoroid is in the Earth's vicinity (several Hill radii, where the Hill 
radius of the Earth $\approx$ 0.01 AU), we use a fixed time step of 1 minute, with the gravitational influence of the Moon being 
separately taken into account. When the meteoroid is sufficiently far away from the Earth, we then increase the time step from 1 
minute to 1 day (from -30 days to -1000 years) in order to speed up the integrations. The orbital elements for each hypothetical 
clone and 2003 EH$_1$ are output at every 10 years.  Furthermore, for all meteoroids, except for EN30109 (the largest at $\approx$ 
1.2 kg.), we examine four different values for the ratio of the solar radiation pressure to the solar gravity "$\beta$ - 
parameter", $\beta=0$, $\beta=10^{-3}$, $\beta=10^{-4}$ and $\beta=10^{-5}$. 

\indent Throughout the backward integrations, we compute the minimum distance between the orbits (MOID) of each clone and the parent 2003 EH$_1$. The aim is to statistically determine the epoch when the MOID between each observed Quadrantid and asteroid 2003 EH$_1$ was at a minimum. We assume that the spreading time in the mean anomaly ($\approx$ 100 years  for JFCs and NEOs \citep{tancredi98}) of meteoroids on their orbit is much shorter than the spreading time in the other orbital elements. Backwards integrations within a few Lyapunov times can reliably provide information on the minimum distance between the orbit of the parent and each Quadrantid, though longer ones may not. Our calculated Lyapunov time for asteroid 2003 EH$_1$ is $\approx$ 80 years which is consistent with the values for JFC and NEOs found by \citet{tancredi98}).  

\indent In addition to the computation of the MOID between the parent and each Quadrantid, we also used the standard orbital 
similarity functions $D_{SH}$ \citep{south} and $D^{\prime}$ \citep{drum}, between the orbits of the parent and each clone as a 
check on our MOID results and found similar behavior. In an attempt to identify a probable formation mode of the core of the 
Quadrantid stream, we also compute the true anomaly $\theta$ at the MOID of the orbit of 2003 EH$_1$ and each clone. Meteoroid 
ejection due to cometary ice sublimation must be confined within an arc of the orbit when the parent is sufficiently close to the 
Sun, i.e within $r=$ 3 AU. The backwards integrations of the equations of motion for the parent, 2003 EH$_1$, indicate that its 
orbital elements do not vary dramatically over 200 years from their present values. If the narrow portion of the Quadrantids was 
created through cometary meteoroid ejection 
around 1800 AD, that must had happened within an arc of $\theta \approx$ $\pm$ 130$^{\circ}$ from perihelion which is roughly equivalent to a heliocentric distance of $r\approx$ 3 AU.

\indent In order to further constrain the meteoroid release mode, we also compute the relative velocity between the parent and each clone at the MOID. The aim is to compare whether the ejection velocities within the arc of water ice sublimation are consistent with cometary dust ejection speeds (see Section~\ref{ejecmodel}). Although, all meteoroid ejection models yield slightly different ejection velocities (a few tens to few hundred m/s), nonetheless they all agree that the latter are unlikely to exceed $\sim$ 1 km/s. Thus, if the relative velocity between the parent, 2003 EH$_1$, and each observed Quadrantid is within few-hundreds m/s, we may argue for cometary origin of the core of the Quadrantid stream.

\subsection{Phase 2: Forward integrations}\label{forward}

Using the results of the backward integrations to provide a time window for the likely formation epoch of the core of the Quadrantid meteoroid stream, we next use ejection models (described below) to generate a hypothetical meteoroid stream and integrate the equations of motion of the meteoroids forward in time. 

\indent Based on the results of Phase 1, we assume that meteoroid ejections took place between 1700 AD and 1900 AD, and we use this time window to eject meteoroids from the parent 2003 EH$_1$ between 1700 - 1900 AD. We explicitly assume that parent, 2003 EH$_1$, was an active comet prior to 1900 AD, supplying meteoroids to the Quadrantid meteoroid stream, and that activity ceased after the beginning of 20$^{th}$ century. The latter assumption will be motivated by our backwards integrations as described later. Thus, we effectively test the hypothesis whether meteoroids ejected during between 1700 - 1900 AD could reproduce the presently observed narrow structure of the Quadrantid stream (see Section~\ref{results} for details).

\indent We first test what is the most likely epoch, at which meteoroids have to be released from 2003 EH$_1$, so the resulting meteoroids start intersecting the Earth's orbit around 1835 AD i.e when the stream was first noticed. For this purpose, we eject 10$^4$ meteoroids, of both radar (100$\mu m$ - 1 mm) and visual sizes (1 mm - 1 cm), at various epochs as a single outburst from the parent and propagated their equations of motion forward in time. The ejection model that we used for the single outburst meteoroid ejection is the one by \citet{BJ98}, described in Section~\ref{ejecmodel}. We note here that the epochs, at which radar and visual size meteoroids are released from the parent so they can reach the Earth around 1835 AD, are slightly different. This is due to the different orbital evolution of micron and centimeter sized meteoroids, for which the the magnitude of the solar radiation pressure force is different. Once the ''correct'' time window of meteoroid release has been identified (i.e the resulting meteoroids reach the Earth around 1835 AD), we then use that time window as a starting point for continues meteoroid ejection over multiple perihelion returns of 2003 EH$_1$. In the case of visual size particles this time window is between 1780 - 1786 AD, whereas for radar size meteoroids it is between 1790 - 1796 AD (see Section~\ref{results}). 

\indent In the case of meteoroid ejections over multiple perihelion returns of the 2003 EH$_1$, 10$^4$ meteoroids are released at each perihelion passage of the parent (within an arc of 3 AU, centered at the perihelion of 2003 EH$_1$), both at radar and at visual sizes, to compare with both observation techniques. The meteoroid ejection speeds considered are described in Section~\ref{ejecmodel}. The number of particles, in a given size bin, were chosen uniformly in the logarithm of the size. During the integrations, the orbital elements, the resultant radiant position and geocentric velocity for the meteoroids are computed for each day. That allows us to closely examine the activity profile of the resultant stream as a function of the meteoroid size, i.e radar and visual meteors, and compare it with the observed activity profile, radiant position and dispersion and geocentric speed. 

\indent We would like to note, that we also performed sample simulations assuming the extreme values for the radius of 2003 EH$_1$, i.e $R_c=$2 km (see Section~\ref{quads&EH1}) as well as the density of the meteoroids within 0.8 g/cm$^3$ $ < \rho < $ 3 g/cm$^3$. However, the variation of the radius of the parent, did not show a noticeable difference in the final results. Furthermore, we also repeated the simulations with all variables as above, i.e parent's radius and density of meteoroids, number of meteoroids ejected per perihelion return of 2003 EH$_1$, meteoroid masses and $\beta$-values etc., thus replacing only the ejection speed model with the one resulting from hydrodynamical study of the cometary circum-nucleus coma by \citep{crifo95, crifo97}. The latter model is generally known to yield slightly lower meteoroid ejection speeds, as compared to e.g \citet{whip51, BJ98}. Effectively, this allowed us to test the significance of the magnitude of the ejection speeds on our final results, keeping all other parameters same as above. However, the overall results did not show a modest difference, thus implying that the considered ejection model has negligible effect on the final results, over the time scale of integrations we are concerned with. We thus, did not see any strong reason to further investigate which ejection model e.g. \citep{BJ98} should be considered over another e.g. \citet{crifo95, crifo97}. Therefore, in this work we only present the results from the simulation, carried out assuming meteoroid ejection speeds modeled by \citep{BJ98}.    

\section{Results}\label{results}

\subsection{Phase 1: Meteoroid release epoch}

\vspace{0.3cm}

In Phase 1, we integrated 10$^4$ hypothetical clones backwards for each observed Quadrantid (eight photographic and five radar). For brevity we show the results, from backward integrations, for only three core and two non-core Quadrantids. Moreover, we show the results for only $\beta = 0$, as the other instances of $\beta$ yielded similar outcomes. The simulations are thus almost insensitive to the meteoroids' mass range, given the time scale of our simulations. Despite solar radiation pressure force being non-negligible for micrometer and millimeter sized particles, we do not find that modestly larger beta values noticeably change the final results.

\begin{figure*}
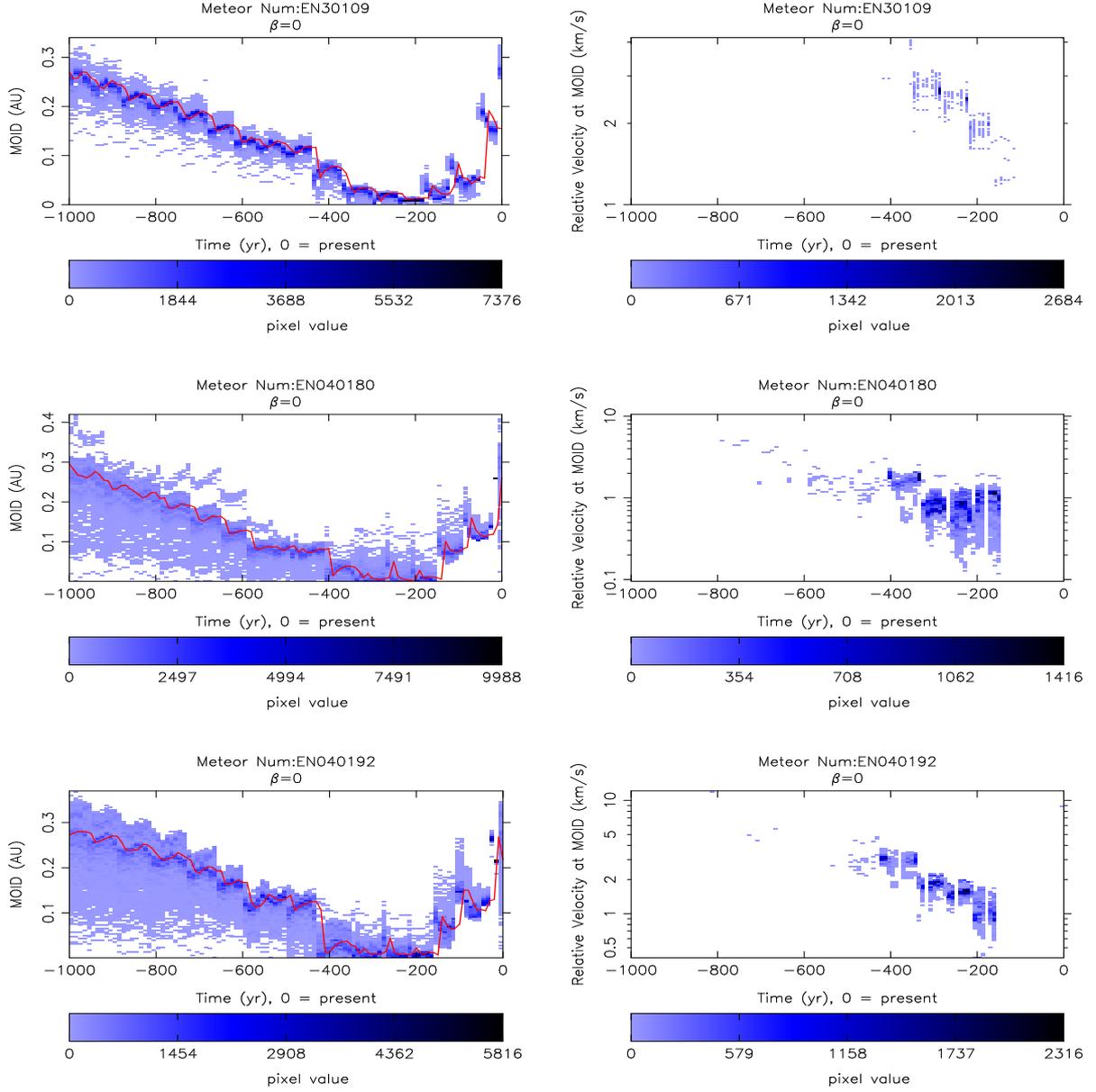
%
\centering
\parbox{7.5cm}{\includegraphics[width=0.65\linewidth,height=7.5cm,angle=-90]{moid-en040192.eps}}%
\qquad
\begin{minipage}{7.5cm}%
\includegraphics[width=0.65\linewidth,height=7.5cm,angle=-90]{vel-040192.ps}
\end{minipage}%
\caption{\footnotesize{The Minimum Intersection Distance (MOID) (left panels) and the relative velocity at the MOID (right panels)  between the orbits of asteroid 2003 EH$_1$ and 10$^4$ clones for each the three photographic meteors, belonging directly to the core of the Quadrantid meteoroid stream. The pixels in the figure are color coded in the hue of the blue color, with a darker blue corresponding to a greater number of clones. The red curve in the right panels correspond to the median value of the MOID at the given epoch. The relative velocity at the MOID is presented in a logarithmic scale in the panels on the right hand side of the figures. Clones with MOID $>$ 0.01 AU are not plotted in the right hand panels.}}%
\label{moid-core}%
\end{figure*}

\indent Fig.~\ref{moid-core} shows the evolution of the MOID (left panels) and the relative velocity at the MOID (right panels), 
between the orbits of 2003 EH$_1$ and 10$^4$ clones, for three core Quadrantids. The y - dimension of each pixel in 
Fig.~\ref{moid-core} is equal to $\sim$ 0.005 AU in the left panels and $\sim$ 0.01 km/s in the right panels, whereas the 
x-dimension is 10 years. It is clearly seen, from Fig.~\ref{moid-core}, that the MOIDs between the clones of each integrated 
core-Quadrantid and 2003 EH$_1$ show a deep minimum $\approx$ 200 - 300 years before the present ($\approx$ 1700 - 1800 AD), with 
little dispersion in the MOID at that particular epoch. The same general tendency of the MOID, with a minimum between 1700 AD 
-1800 AD, was observed even for particles with extreme values of $\beta = 10^{-3}$.

\begin{figure*}%
\centering
\parbox{7.5cm}{\includegraphics[width=0.65\linewidth,height=7.5cm,angle=-90]{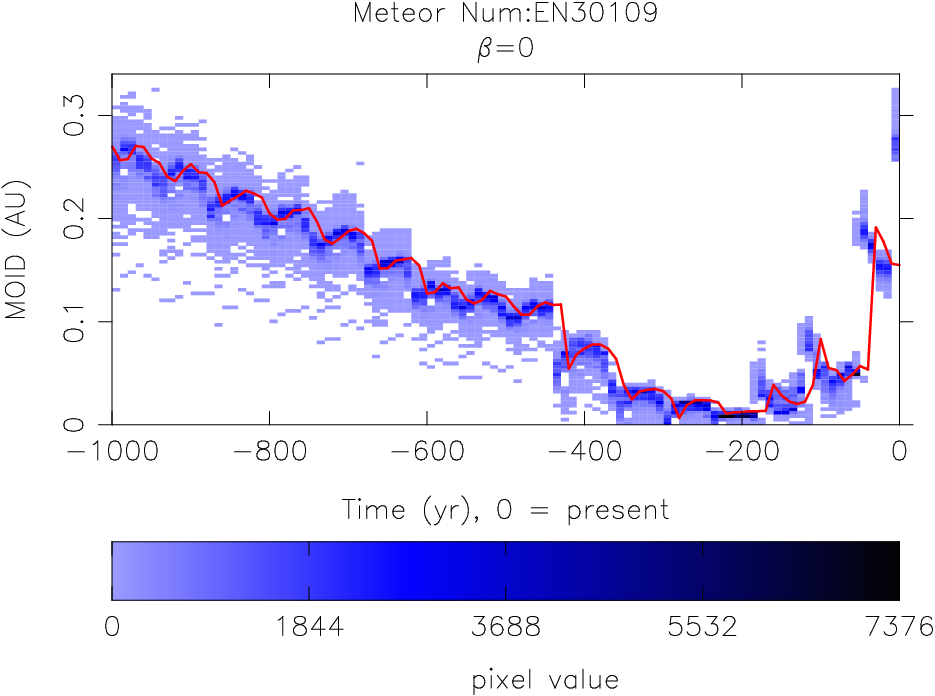}}%
\qquad
\begin{minipage}{7.5cm}%
\includegraphics[width=0.65\linewidth,height=7.5cm,angle=-90,]{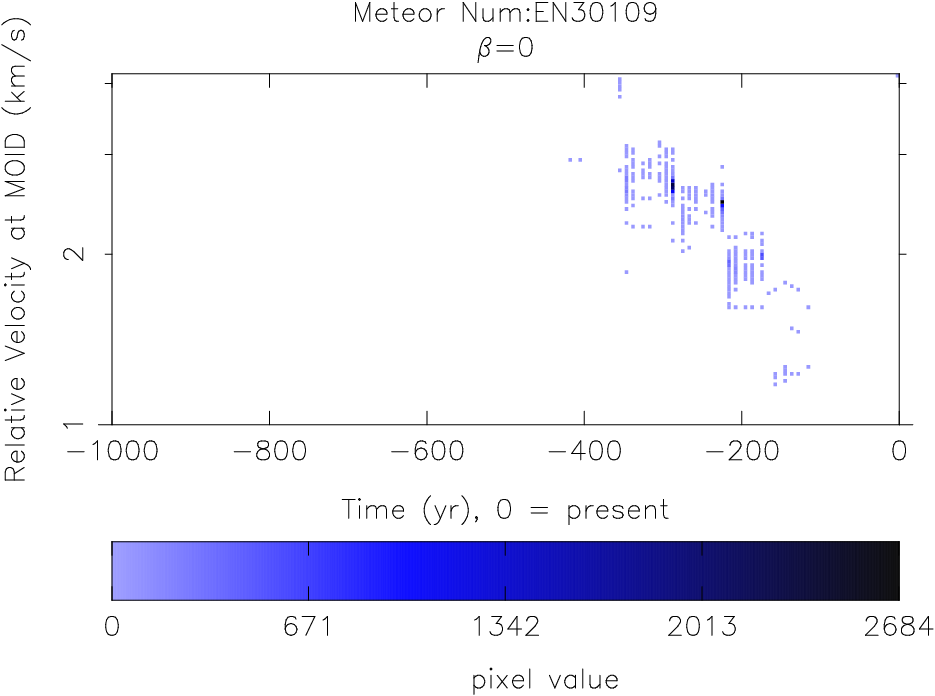}
\end{minipage}%
\label{fig:1figs}%
\end{figure*}

\begin{figure*}%
\centering
\parbox{7.5cm}{\includegraphics[width=0.65\linewidth,height=7.5cm,angle=-90]{moid-en040180.eps}}%
\qquad
\begin{minipage}{7.5cm}%
\includegraphics[width=0.65\linewidth,height=7.5cm,angle=-90,]{vel-040180.ps}
\end{minipage}%
\label{fig:1figs}%
\end{figure*}

\begin{figure*}
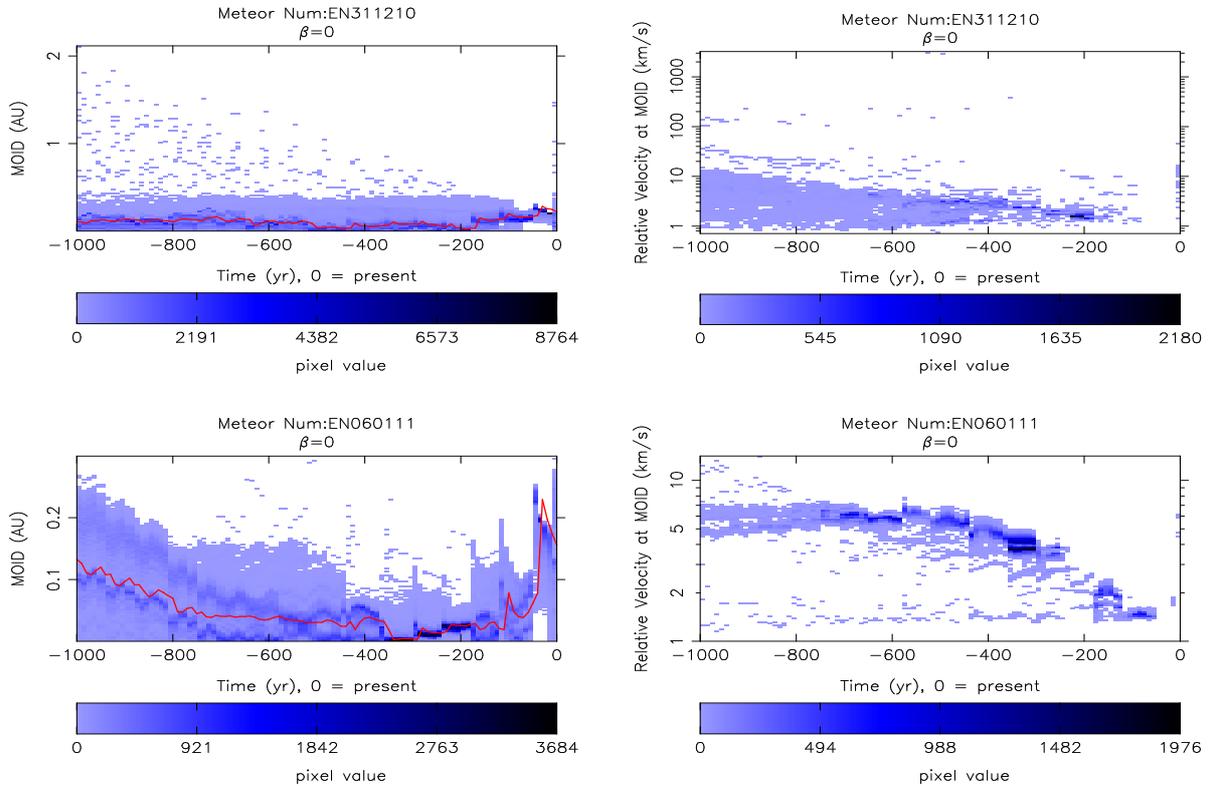
%
\centering
\parbox{7.5cm}{\includegraphics[width=0.65\linewidth,height=7.5cm,angle=-90]{moid-en060111.eps}}%
\qquad
\begin{minipage}{7.5cm}%
\includegraphics[width=0.65\linewidth,height=7.5cm,angle=-90]{vel-060211.ps}
\end{minipage}%
\caption{\footnotesize{The Minimum Intersection Distance (MOID) (left panels) and the relative velocity at the MOID (right panels)  between the orbits of asteroid 2003 EH$_1$ and 10$^4$ clones for two photographic meteors, not belonging directly to the narrow core of the Quadrantids. Clones with MOID $>$ 0.01 AU are not plotted in the right hand panels.}}
\label{moid-noncore}%
\end{figure*}

\indent Similarly, Fig.~\ref{moid-noncore} shows the evolution of the MOID (left panel) and relative velocity at the MOID (right panel), as above, for "non-core" (observed outside the narrow peak of the stream) meteoroids. Unlike core-Quadrantids, their non-core counterparts did not show an obvious minimum of the MOID. The exception was non-core meteor number EN060111 which did show a weak minimum in the MOID around 1650 AD but with a relatively high dispersion. We suggest that the non-core Quadrantids are much older, though the Lyapunov times are such that longer backwards integrations to confirm this hypothesis are problematic.

\begin{figure*}
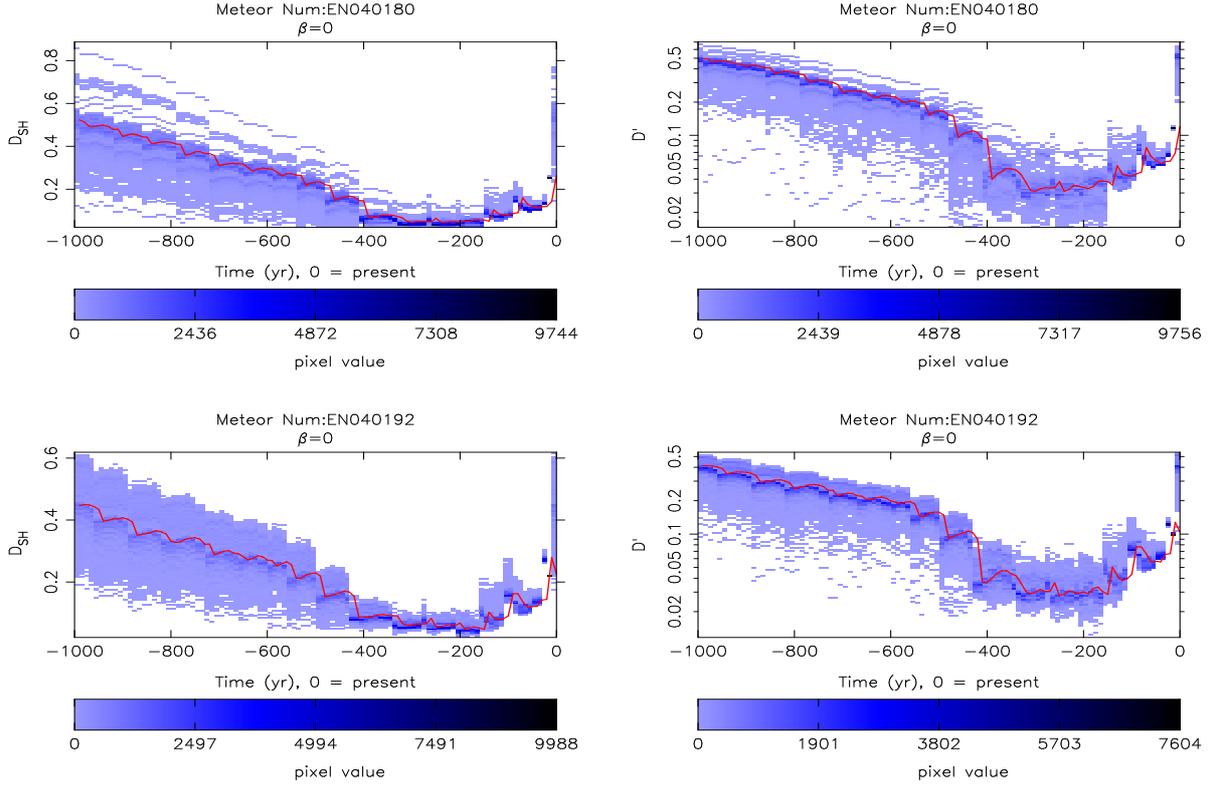
%
\centering
\parbox{7.5cm}{\includegraphics[width=0.65\linewidth,height=7.5cm,angle=-90]{en040192-dsh.eps}}%
\qquad
\begin{minipage}{7.5cm}%
\includegraphics[width=0.65\linewidth,height=7.5cm,angle=-90]{en040192-dcrt.eps}
\end{minipage}%
\caption{\footnotesize{The evolution of the similarity criteria, $D_{SH}$ \citep{south} (left panel) and $D^{\prime}$ 
\citep{drum} (right panel), for 10$^4$ clones of each Quadrantid as a function of time. The red curve corresponds to the median 
value of the similarity functions, $D_{SH}$ and $D^{\prime}$, respectively.}}%
\label{dcrits}%
\end{figure*}

\indent As an additional test as to the age of the core of the Quadrantid stream, we also computed the standard similarity 
functions $D_{SH}$ \citep{south} and $D^{\prime}$ \citep{drum}, between the orbits of the clones of each individually observed 
core-Quadrantids and 2003 EH$_1$ Fig.~\ref{dcrits}. However, here we present the results for only 3 bolides, EN030109, EN040180 
and EN040192. The results for the rest of the core-Quadrantids yielded similar results. The size of each pixel in the left panels 
is $\delta x=10$ years, $\delta y=0.01$, whereas in the right panels ($D^{\prime}$ criterion) $\delta x=10$ years, $\delta 
y=0.005$. Both similarity criteria yielded an unambiguous deep minimum between 1700 - 1800 AD, similar to the MOID criterion, for 
the core-Quadrantids. However, in the case of non-core Quadrantids there was not a clear minimum of the similarity function, the 
results which we have omitted here.   

\indent The effect of varying the $\beta$ - parameter had little effect, that is the evolution of both, $D_{SH}$ and 
$D^{\prime}$ criteria, yielded a minimum between 1700 - 1800 AD. The latter result is not surprising, given the time scale of our 
simulations and in particular the time of interest (200 - 300 years), i.e a few Lyapunov times.   

\begin{figure*}%
\centering
\parbox{7.5cm}{\includegraphics[width=0.65\linewidth,height=7.5cm,angle=-90]{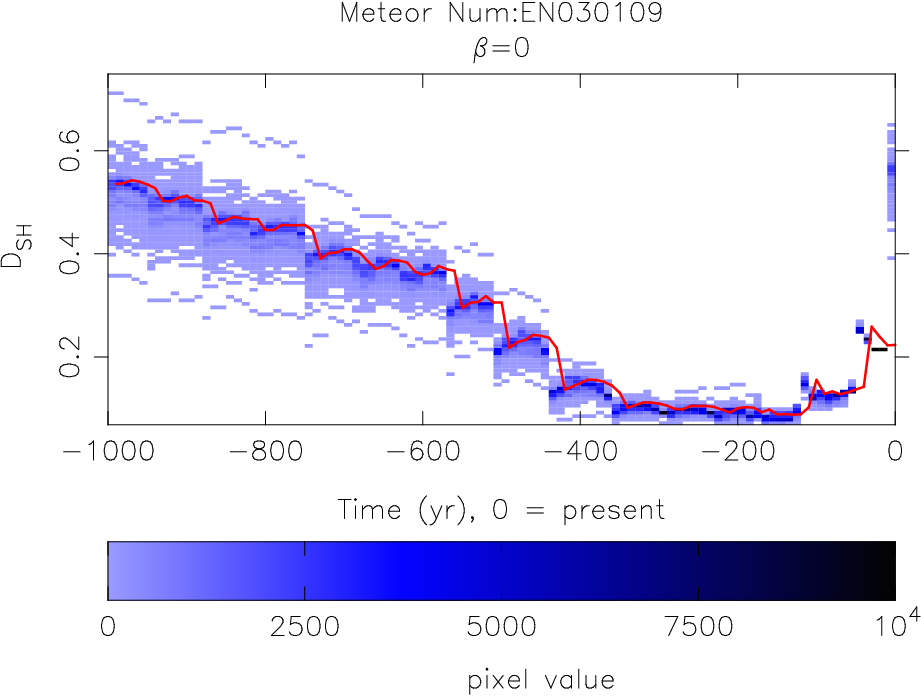}}%
\qquad
\begin{minipage}{7.5cm}%
\includegraphics[width=0.65\linewidth,height=7.5cm,angle=-90,]{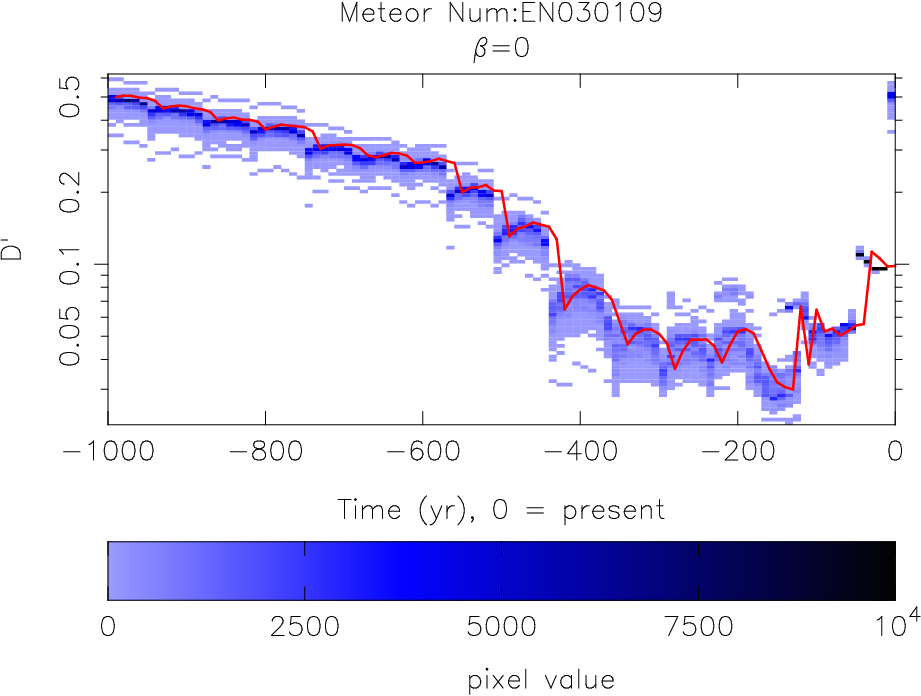}
\end{minipage}%
\label{fig:1figs}%
\end{figure*}

\begin{figure*}%
\centering
\parbox{7.5cm}{\includegraphics[width=0.65\linewidth,height=7.5cm,angle=-90]{en040180-dsh.eps}}%
\qquad
\begin{minipage}{7.5cm}%
\includegraphics[width=0.65\linewidth,height=7.5cm,angle=-90,]{en040180-dcrt.eps}
\end{minipage}%
\label{fig:1figs}%
\end{figure*}

\begin{figure*}%
\centering
\parbox{7.5cm}{\includegraphics[width=0.65\linewidth,height=7.5cm,angle=-90]{moid-en311210.eps}}%
\qquad
\begin{minipage}{7.5cm}%
\includegraphics[width=0.65\linewidth,height=7.5cm,angle=-90,]{vel-311210.ps}
\end{minipage}%
\label{fig:5figs}%
\end{figure*}

\begin{figure*}
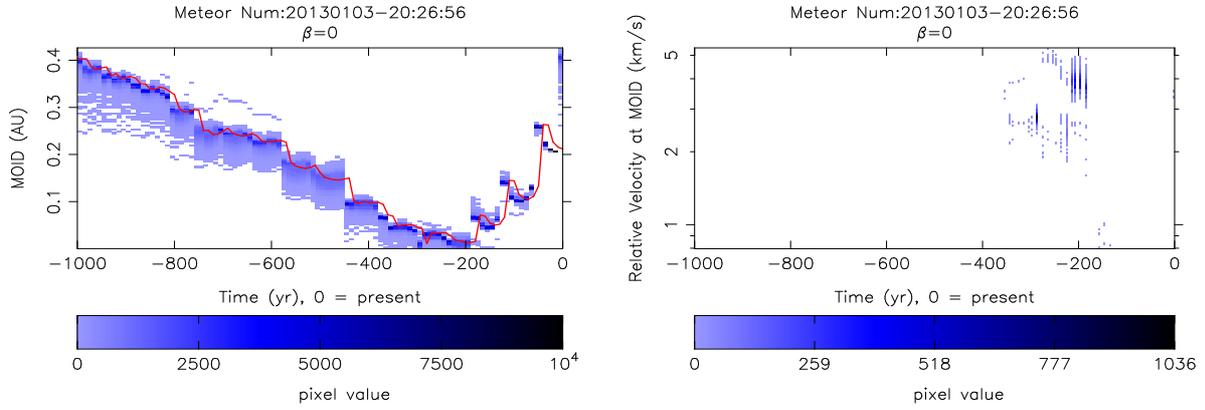
%
\centering
\parbox{3.2in}{\includegraphics[width=0.65\linewidth,height=7.5cm,angle=-90]{fig.eps}}%
\begin{minipage}{3.2in}%
\includegraphics[width=0.65\linewidth,height=7.5cm,angle=-90]{vel.eps}
\end{minipage}%
\caption{\footnotesize{Evolution of the Minimum Orbit Intersection Distance (MOID), between the orbit of 2003 EH$_1$ and ten-thousand fictitious clones of the radar meteor - 20130103-20:26:56 over a period of one-thousand years. This radar meteor belongs to the core of the Quadrantids. The red curve represents the median value of the MOID at a given epoch. Clones with MOIDs $>$ 0.01 AU are not plotted in the right hand panels.}}
\label{moid-radar}
\end{figure*}

\begin{figure*}
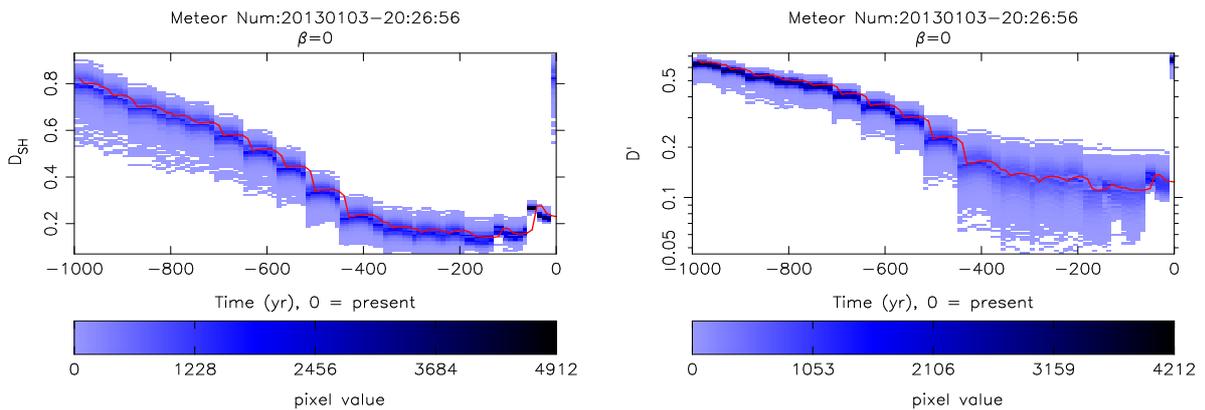
%
\centering
\parbox{3.2in}{\includegraphics[width=0.65\linewidth,height=7.5cm,angle=-90]{radar-dsh.eps}}%
\begin{minipage}{3.2in}%
\includegraphics[width=0.65\linewidth,height=7.5cm,angle=-90]{radar-dcrt.eps}
\end{minipage}%
\caption{\footnotesize{Evolution of the orbital similarity criteria, $D_{SH}$ and $D^{\prime}$, between the orbit of 2003 EH$_1$ 
and ten-thousand fictitious clones of the radar meteor - 20130103-20:26:56 over a period of one-thousand years. This radar meteor 
belongs to the core of the Quadrantids. The red curve represents the median value of the similarity functions.}}
\label{dcrits-radar}
\end{figure*}

\indent We present the results for only one radar core-Quadrantid, since all other cases yielded similar results. From 
Fig.~\ref{moid-radar} it can be seen that, as in the case of the photographic core-Quadrantids, the minimum value of the MOID 
between the orbits of the parent and the fictitious clones of meteor 20130103-20:26:56 is reached around 1700 - 1800 AD. The 
simulations for different $\beta$ - values also yield similar results. Fig.~\ref{dcrits-radar} shows the evolution of the orbital 
similarity functions, $D_{SH}$ and $D^{\prime}$, between 10$^4$ clones of the radar core Quadrantid 20130103-20:26:56 and 2003 
EH$_1$. In this case, the minimum of the similarity criteria was somewhat less obvious but still around 1800 AD. We note that 
radar orbital measurements are of lower precision compared to  photographic or video observations. Nevertheless, the minima of 
the similarity functions, occurring after 1700 AD supports the hypothesized young age of the core of the Quadrantids.

\indent As was the case with the photographic core Quadrantids, variation of the ratio of the solar radiation pressure to the solar gravity $\beta$ did not change the overall results for radar meteoroids as to the position of the minima of both similarity functions. The latter result seems to be reasonable, given the time window we are interested in (200-300 years). 

\indent All lines of evidence from the backward integrations in Phase 1 point strongly to an origin in the past 200-300 years for the core Quadrantids.

\subsection{Phase 1: Formation mechanism of the Quadrantid meteoroid stream}

\vspace{5mm}

\indent Another goal of this work is to constrain the formation mode of the core of the Quadrantid meteoroid stream. More precisely, we aim to test if it is consistent with the hypothesis of cometary origin.
 
\indent We recall that existing dust ejection models from comet nuclei suggest that the ejection velocities are much smaller (few tens-hundreds m/s) than the orbital speed of the comet (few tens of km/s at perihelion), with extreme ejection velocities not exceeding 1 km/s (see e.g., \citep{whip50,whip51,jones95,crifo95,crifo97}).

\indent In order to test if the Quadrantid meteoroid stream may have resulted from a cometary activity on 2003 EH$_1$ we follow an approach similar to \citet{gust89} and \citet{adol96}, comparing the relative velocity at the MOID between 10$^4$ clones of each photographic and radar core-Quadrantid and 2003~EH$_1$. It has to be noted, though, that this approach has its limitations and should be used carefully. The uncertainties in all measurements e.g. position of the meteoroids and velocities are finite which over time will tend to increase. Therefore we emphasize that, we utilize this method assuming the age of the ejected meteoroids is less that a few Lyapunov spreading times of the meteoroids along their orbits. Secondly, there are uncertainties in the calculations of the meteoroids' initial (pre-atmospheric) masses, where the latter come into the equations of motion primarily by their $\beta$- values. Thus, the results from the simulations must be treated from a statistical point of view. Nevertheless, assuming that the core of the Quadrantid meteoroid stream is relatively young , compared to the Lyapunov spreading time, we should be able to approximately constrain its age, treating the results from the backward and forward integrations in a statistical sense. 

\indent The results for the relative velocity between 2003 EH$_1$ and $10^4$ clones of each photographic Quadrantid is presented in Fig.~\ref{moid-core} (right panels) and radar Quadrantids in Fig.~\ref{moid-radar} (right panels). The epoch of interest is the one centered between -200 and - 300 years from the present. Relative velocities are computed only for  clones with MOIDs less than 0.01~AU from the orbit of 2003~EH$_1$. 

\indent During the minimum of the MOID (left panel), the relative velocity at the MOID between the clones of each core Quadrantid and the parent 2003 EH$_1$, range from about 0.1 to 2 km/s.  These values are on the high side for cometary ejection processes, as the ejection speeds of meteoroids leaving the surface of 2003 EH$_1$, given the perihelion distance, even for micrometer meteoroids are unlikely to exceed 200 m/s. However, the process of generating clones within the uncertainties necessarily generates many incorrect orbits: only one clone within the ensemble is the real particle. Thus, the existence of even one clone with a relative speed at the MOID suggests a possible cometary origin.

\indent The lowest relative velocities within the clones of the core Quadrantids and 2003 EH$_1$, range from 200 m/s - 800 m/s, with the majority of the clones having relative velocities exceeding 1 km/s. Only 2 clones of bolide EN040180 attained a very low relative speed of $\approx$ 80 m/s while most had speeds as great as a few km/s. We conclude that these results are marginally consistent with cometary ejection processes but can not be used as definitive proof for a cometary sublimation origin for the core of the stream.

\indent Similar results are found for the radar core Quadrantids (Fig.~\ref{moid-radar}).  Here we see even larger dispersion in the ejection speed from the parent, ranging from 0.8 km/s to $\sim$ 5 km/s, around 1800 AD. However, only between 1 - 4 particles demonstrated relative speeds of 0.8 km/s, while the majority had speeds above 1 - 3 km/s. This result is poorly consistent with the expected meteoroid ejection speed from comets, i.e it is unlikely that meteoroids would leave the surface of the parent with speeds of 800 m/s, at a heliocentric distance of 1 AU under cometary sublimation alone. We recall, however, that the errors in the initial values of the position and speed of the radar meteoroids in the Earth's atmosphere are larger than those obtained by photographic observations.

\indent As an additional constraint, as to the more likely formation mechanism of the core of the Quadrantids, we also calculated the true anomaly at the MOID, of the orbit of 2003 EH$_1$, for 10$^4$ clones of each Quadrantid (see Section~\ref{backward}). Sublimation of cometary ices must take place closer to the Sun (within $\sim$ 3 AU for 2003 EH$_1$, assuming sublimation of water ice only).

\begin{figure}[h!]
\begin{center}
\advance\leftskip -3cm
\advance\rightskip -3cm
\includegraphics[width=0.7\linewidth,angle=-90]{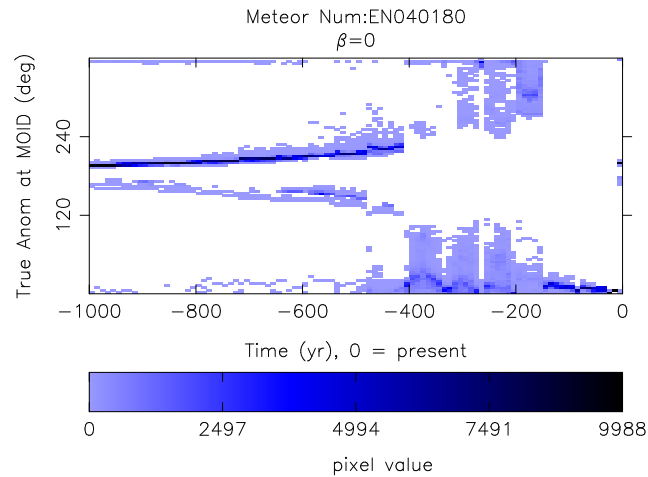}
\caption{\footnotesize{The true anomaly of 2003 EH$_1$ at the MOID of 10$^4$ fictitious meteoroids (clones) of the core-Quadrantid photographic meteor - EN040180 for a time span of 1000 years.}}
\label{truan}
\end{center}
\end{figure}

\begin{figure}[h!]
\begin{center}
\advance\leftskip -3cm
\advance\rightskip -3cm
\includegraphics[width=0.7\linewidth,angle=-90]{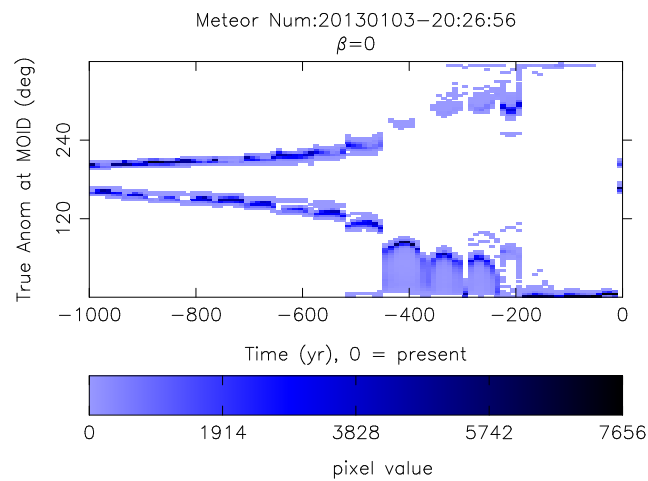}
\caption{\footnotesize{The true anomaly of 2003 EH$_1$ at the MOID of 10$^4$ fictitious meteoroids (clones) of the radar meteor - 20130103-20:26:56, for a time span of 1000 years.}}
\label{truan-radar}
\end{center}
\end{figure}

Fig.~\ref{truan} and~\ref{truan-radar} shows the true anomaly at the MOID for ten-thousand clones as a function of time of the photographic meteoroid EN040180 (1.9 g) and radar event 20130103-20:26:56, both of which are core-Quadrantids. It can be seen that, for the epoch 1800 AD, there is a greater probability that the true anomaly at the MOID, on the orbit of 2003 EH$_1$, was between $\nu \approx \pm 60^{\circ}$ from perihelion (translating to a heliocentric distance $r \approx$1.3 AU), which is consistent with cometary activity, resulting in meteoroids ejection from 2003 EH$_1$. 

\indent Examination of the results for the rest of the photographic and radar meteors in our sample were similar to those in Fig.~\ref{truan} and~\ref{truan-radar}, with some showing lower or higher dispersion in true anomaly at the MOID at the epoch of interest (1700 - 1900 AD). The effect of varying the $\beta$ - value had little effect on the final outcome. The highest median value of the true anomaly at the MOID was $\nu \approx \pm 120^{\circ}$ from perihelion (roughly corresponding to a heliocentric distance $r\approx$~2.4 AU). Nonetheless, all results showed a higher probability that the true anomaly at the MOID on 2003 EH$_1$ occurred close to perihelion, consistent with meteoroid ejection due to sublimation of ices.

\indent The location of the MOIDs are certainly consistent with cometary ejection processes, while the relative velocities are low enough in some cases, but not all. Since, it is not possible to know the "true" backward evolution of any of the observed Quadrantid meteoroids, due to the inherent uncertainties in their position and speed, thus if even one clone has relative speed consistent with cometary ejection, it may point to origin consistent with cometary sublimation.     
We conclude here that cometary ejection processes may be a likely source of the core Quadrantids, or at least that we have sufficient grounds to proceed with forward modeling of the stream under this assumption.

\subsection{Phase 2: Meteoroid ejection}

\indent In order to test the results obtained by backwards integrations, we employ the "direct approach" by ejection of a large number of hypothetical meteoroids from the parent, and integrate them forward in time. By the "direct approach" we test if the observed average physical characteristics of the Quadrantid stream, described in Section~\ref{intro} and~\ref{quads&EH1}, are consistent with cometary ejection from 2003 EH$_1$ around 1700 - 1900 AD. 

\indent We briefly recall the constraints that we are attempting to match by the forward integrations.

1. The timing of the appearance of the stream on the sky (around 1835 AD) \citep{quet}.

2. The mean position and spread of the geocentric radiant of the stream (e.g. \citep{jen97}).

3. The position of the peak of the activity profile of the core (e.g. \citep{Ren93}).

4. The width of the activity profile of the core (FWHM $\approx$ 0.6 days) \citep{hugh77, Ren93, jen06}.

\indent We first show the results for $10^4$ meteoroids, ejected from 2003 EH$_1$, at different epochs in a single outburst (at a single point on the orbit). The aim is to obtain a time windows as to when meteoroids must be ejected so they can reach the Earth's orbit around 1835 (see Section~\ref{forward}). Thus, meteoroids ejected prior to that time window will reach the Earth before 1835, i.e when the shower was first observed. Furthermore, we have to identify the epoch of ejection when meteoroids cease reaching the Earth. We then use these time windows for meteoroid ejection, at every perihelion passage in order to closely inspect the characteristics of these synthetic meteor showers.

\begin{figure*}[!htp]
\centering
\subfigure[]{
\includegraphics[width=0.47\textwidth]{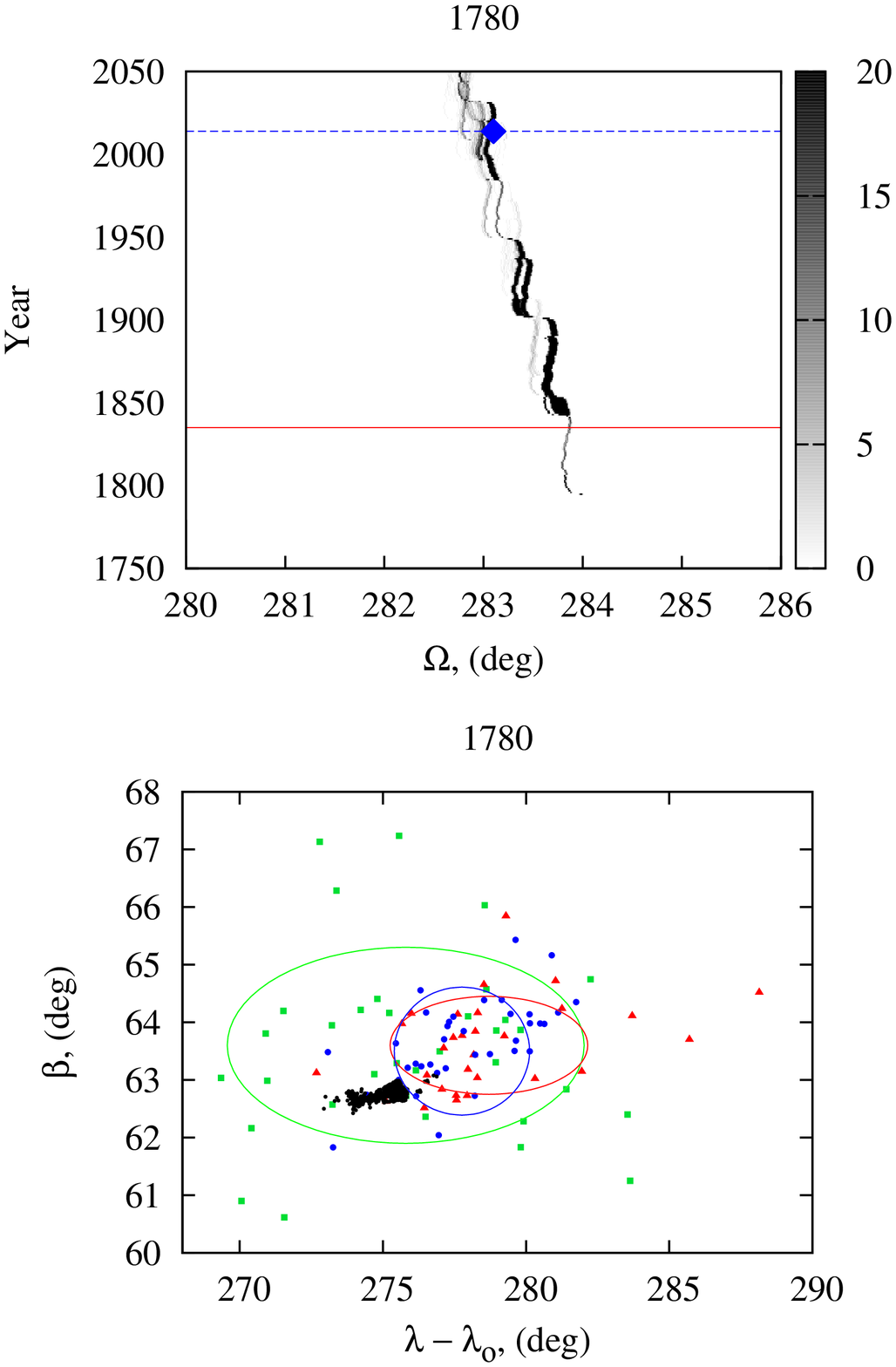}
}

\subfigure[]{
\includegraphics[width=0.47\textwidth]{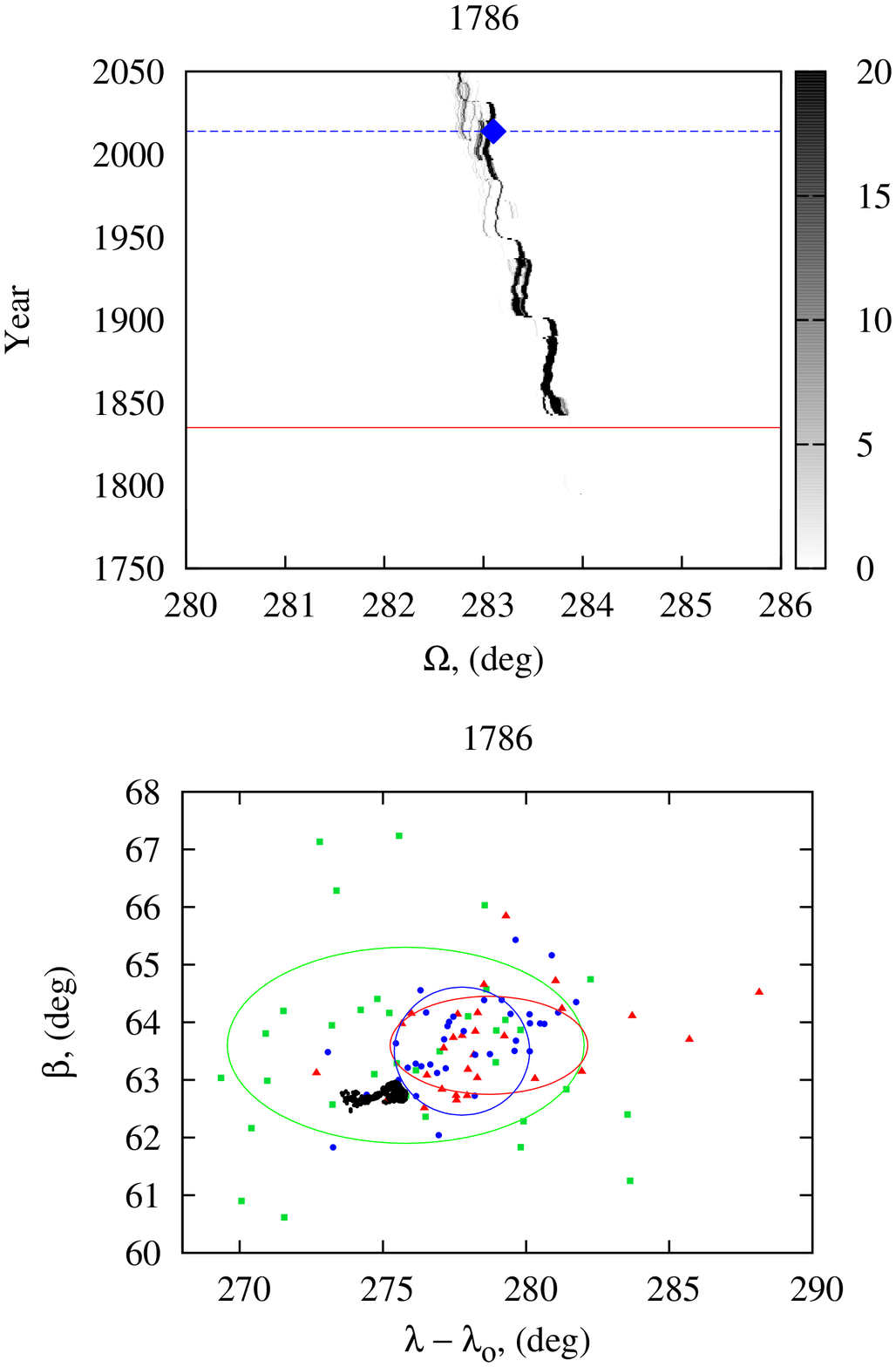}
}
\caption{\footnotesize{Evolution of the descending nodes of meteoroids (upper panels) ejected from 2003 EH$_1$ in 1780 and 1786. The red line marks the year of 1835 AD, i.e the first reported appearance of the Quadrantids \citep{quet}. The blue line marks the year 2014, whereas the blue diamond corresponds to the solar longitude of the peak of the shower. Only meteoroids with nodes within 0.01~AU of the Earth's orbit are plotted. \\
\indent The lower panels show the theoretical geocentric radiant position of meteoroids (black dots), ejected from 2003 EH$_1$ at the given ejection epoch, superimposed over the observed mean radiant positions measured by video and photographic observations. The red triangles are individual Quadrantid radiant positions from the {\it{Dutch Meteor Society}} DMS - video data \citep{Bet95}, the blue circles from the DMS - photographic data and the green squares are the Quadrantid radiants as detected by the {\it{Southern Ontario Meteor Network}} (SOMN) all-sky video systems \citep{brown-weryk10}. The corresponding ellipses are centered at the mean values of the individual data sets, whereas the ellipses denote the dispersions of the corresponding techniques (see Table~\ref{radiant-coords}).}}
\label{vis-nd-rad}
\end{figure*}

\indent Fig.~\ref{vis-nd-rad} (upper panels) shows the evolution of the descending nodes of visual size meteoroids ejected from 2003 EH$_1$, in a single perihelion passage, in 1780 and 1786, respectively. Also, Fig.~\ref{vis-nd-rad} (lower panels) show the mean radiant position, of the simulated meteoroids, in sun-centered ecliptic longitude $\lambda - \lambda_{\odot}$ and latitude $\beta$. In the figures, only meteoroids having their descending node within 0.01 AU, from the orbit of the Earth are plotted so we can sample only those particle which may intersect the Earth. It is evident that meteoroids ejected prior to 1780 arrive near the Earth's orbit somewhat before 1835 (see Fig.~\ref{vis-nd-rad}-a), which is inconsistent with the timing of the shower's appearance (circa 1835 AD). However, it is difficult to argue when exactly meteoroids must had been released from the parent in order to arrive at the Earth at the right time (circa 1835). This is partly due to the unrealistically low number of the simulated particles considered in our simulations, as well 
as the lack of exact conditions during meteoroid ejection. In fact, in Fig.~\ref{vis-nd-rad}-a, although meteoroids ejected in 1780 arrive at the Earth a few decades before the shower's first appearance in the sky (1835), the activity level could had been much lower and the shower could had easily gone undetected. Therefore, we choose to use as initial ejection epoch the year 1780 instead of 1786.

\indent The theoretical radiant position of the meteoroids (Fig.~\ref{vis-nd-rad}) is plotted along with the mean observed positions and dispersions of the Quadrantids as obtained by photographic and video meteor surveys. The SOMN Quadrantids were selected as being potential Quadrantids if their radiant was within 5 degrees of the nominal Quadrantid radiant and if their speed was within 10\% of nominal Quadrantid speed. Similarly, Fig.~\ref{vis-nd-rad}-b shows that meteoroids ejected around 1786 appear at the Earth just slightly after 1835 AD. The slight timing inconsistency is perhaps due to small number statistics. In fact, out of 10$^4$ particles ejected, only about 30 reach the Earth. Nevertheless, within the statistics considered in our simulations, the time of ejection and first appearance of the stream on the sky is in a good agreement with the minimum of the MOID in the backward integrations (Section~\ref{backward}). The theoretical radiant position seems to be in a reasonable agreement with the observations by SOMN, although there is slight discrepancy with the DMS photographic and video data (Fig.~\ref{vis-nd-rad}-a, b lower panel). Thus particles ejected from the parent in 1786 by cometary processes produce a reasonable match with the observed shower properties. However, the descending nodes of meteoroids ejected after 1882 AD start to gradually move outward from the Earth's orbit, and eventually stop intersecting the Earth. The reason for that is the secular precession of the orbital elements, causing the heliocentric distance of the descending nodes of the meteoroids to recede slowly from the orbit of the Earth. Thus the current visual activity of the core of the Quadrantids shower can be explained by cometary activity of 2003~EH$_1$ between the years 1786 and 1882. The parent cannot have been active earlier because the Quadrantids shower would have become active earlier than is recorded in historical records. Whether the parent may have been active after 1882 remains unconstrained, except by the fact that dust production must have eventually diminished to the currently observed inactive state.

\begin{table}[!htpb]
 \centering
{\footnotesize
\begin{tabularx}{\linewidth}{lXXXX}
\toprule[1pt]
Source               & $\lambda$ - $\lambda_{\odot}$&   $\beta$&  $\sigma$ ($\lambda$ - $\lambda_{\odot}$)& $\sigma$ $\beta$\\
  & (deg)& (deg)& (deg)& (deg) \\\hline
                
Photographic &  &  &  & \\
---DMS               & 277.8& 63.5& 2.4& 1.1\\\\

Video        &  &  &  & \\
---DMS               & 278.7 & 63.6 & 3.5 & 0.9\\
---SOMN            & 275.8 & 63.6 & 6.2  & 1.7\\\\

Radar        &  &  &  & \\
---CMOR     & 277.2 & 63.2 & 6.8 & 3.1 \\        
\bottomrule
\end{tabularx}
}
\hfill{}
\caption{\footnotesize{Radiant position and dispersion of the Quadrantid meteor shower obtained by photographic, video and radar techniques. The coordinates of the average radiant are given in the sun-centered ecliptic frame - ecliptic longitude ($\lambda - \lambda_{\odot}$) and ecliptic latitude $\beta$.
}}
\label{radiant-coords}
\end{table}

\begin{figure*}[!htpb]
\centering
\subfigure[]{
\includegraphics[width=0.47\textwidth]{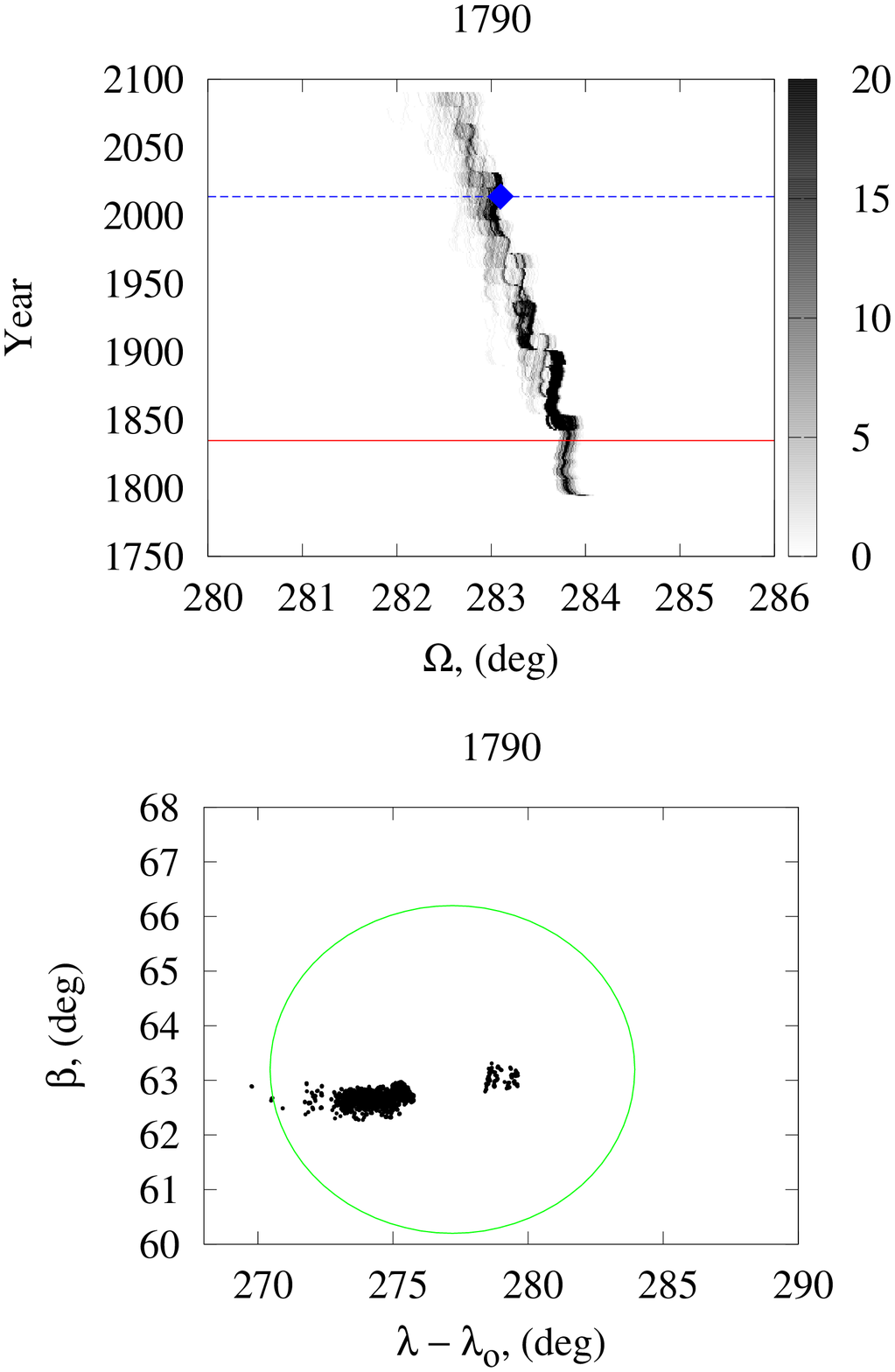}
}
\subfigure[]{
\includegraphics[width=0.47\textwidth]{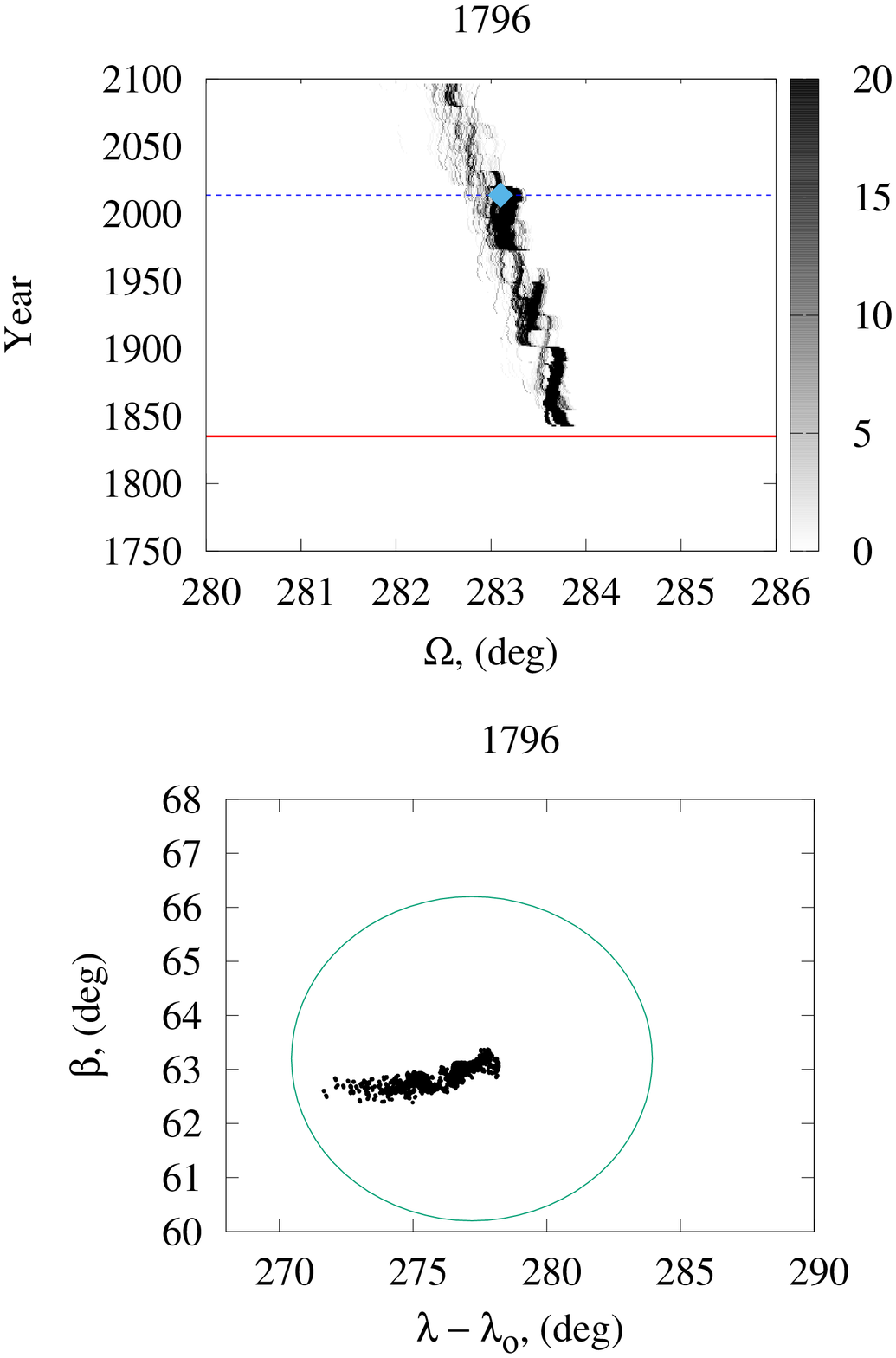}
}
\caption{\footnotesize{Evolution of the descending nodes of radar size meteoroids (upper panels), ejected from 2003 EH$_1$ in 1790 and 1796, until the present time. The red line marks the year of 1835 AD. The blue line marks the year of 2014, whereas the blue diamond corresponds to the solar longitude of the peak of the radar shower. \\
\indent The lower panels show the simulated radiant positions (black dots). The green ellipse is centered at the mean radiant position as deduced from radar observations (CMOR) (see Table~\ref{radiant-coords}. Here the CMOR radiant is that deduced from a 3D wavelet transform isolating the location of the peak average radiant between 2012 - 2014 using the same technique as described in \citet{brown10}. In both figures, only simulated meteoroids approaching the Earth's orbit within 0.01 AU are plotted.}}
\label{radar-nd-rad}
\end{figure*}

\indent Similarly, Fig.~\ref{radar-nd-rad} shows the evolution of the descending nodes of radar size (100 $\mu m$ - 1 mm) meteoroids, ejected from 2003 EH$_1$ in 1790 AD and 1796 AD. Meteoroids, ejected prior to 1796, appear at the Earth as early as 1800 AD. However, we do not really know when radar size particles first reached the Earth, as the earliest Quadrantid meteor radar observations date back only about six decades. The radiant position Fig.~\ref{radar-nd-rad}-a (lower panel) seems to fit well the observed mean Quadrantid radiant as measured by the CMOR. Thus 1796 AD seems to be in a good agreement with the minimum of the MOID, between the orbits of radar meteoroids and that of 2003 EH$_1$ (Section~\ref{backward}). Meteoroids ejected after 1886 AD, from 2003 EH$_1$, do not presently intersect the Earth's orbit due to, as in the case of visual meteoroids, secular precession of the orbital elements of the meteoroids. 

\indent We next present the results of meteoroids ejected over multiple perihelion returns of 2003 EH$_1$ for both, visual and radar size particles. According to our analysis from meteoroid ejections at a single point on the orbit of the parent, we obtained that visual size meteoroids appear and intersect the Earth if ejected roughly between 1780 - 1882 AD, whereas for radar size meteoroids this time window is 1796 - 1886. In this context, approximately 10$^4$ meteoroids were ejected per perihelion passage of the parent, within an arc of 3 AU from the Sun. In the case of the visual size particles, the meteoroids were ejected over  $\sim$ 18 perihelion passages of 2003 EH$_1$ with total number of ejected particles $\approx$ 1.8$\times 10^5$, whereas in the case of radar size meteoroids the the equivalent orbital revolutions of 2003 EH$_1$ are 16, totaling in 1.6$\times 10^5$ ejected meteoroids.

\begin{figure}[!htp]
\begin{center}
\advance\leftskip -3cm
\advance\rightskip -3cm
\includegraphics[width=0.75\linewidth,angle=-90]{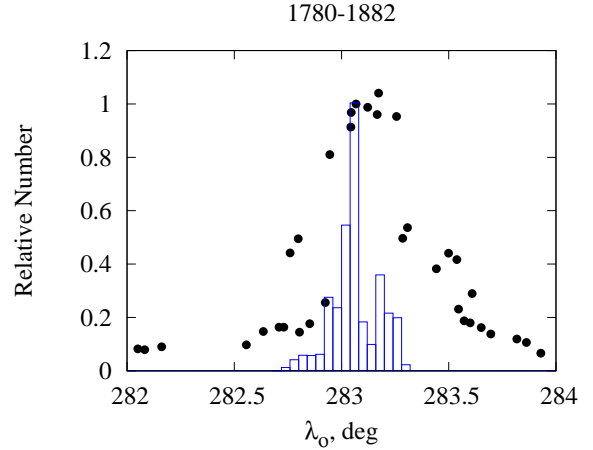}
\caption{\footnotesize{The simulated visual activity profile (blue boxes), superimposed over the observed activity profile (black dots) of the Quadrantid shower. Meteoroids not approaching the Earth's orbit within 0.01 AU are not plotted. The theoretical profile is a stack of 17 perihelion passages of 2003 EH$_1$, corresponding to meteoroid ejections from 1780 - 1882 AD. The observed activity profile is an average activity profile of a few years between 1986 - 1992 (J2000) \citep{Ren93}.}}
\label{actprof-vis}
\end{center}
\end{figure}

\begin{figure}[!htp]
\begin{center}
\advance\leftskip -3cm
\advance\rightskip -3cm
\includegraphics[width=0.8\linewidth,angle=-90]{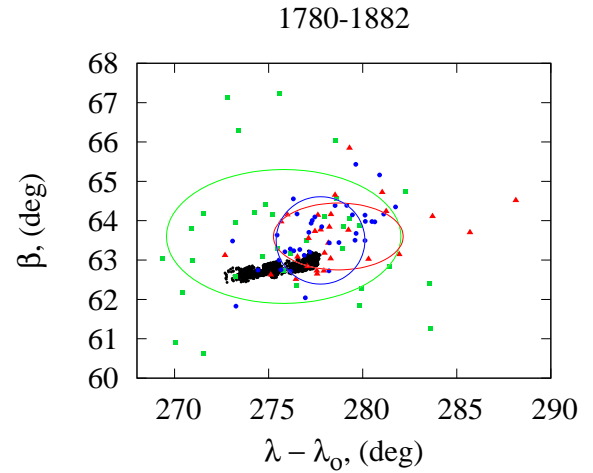}
\caption{\footnotesize{The simulated radiant positions of the visual Quadrantids (black dots), along with the observed mean radiant positions and dispersions. The average radiant position is a stack of 17 perihelion passages of 2003 EH$_1$, corresponding to meteoroid ejections from 1780 - 1882 AD. Meteoroids which do not approach the Earth's orbit within 0.01 AU are not plotted. The observed individual radiants and colors are the same as in Fig.~\ref{vis-nd-rad}.}}
\label{radiant-vis}
\end{center}
\end{figure}

\indent The stacked  theoretical activity profile (blue boxes) of the visual Quadrantids ejected between 1780 and 1882 AD, is presented in Fig~\ref{actprof-vis}. It can be seen that the location of the peak activity of our simulated Quadrantid stream, as well as the FWHM of the core activity, match fairly well with the observations. The wider portion of the observed activity profile though is not well reproduced; we suspect these wings are likely much older than the timescales (of order a few hundred years) that we are concerned with it in this work. The theoretical radiant of meteoroids, ejected between 1780 and 1882 AD, shows a fairly good match with the observed mean position and dispersion of individual visual Quadrantids (Fig.~\ref{radiant-vis}).

\begin{figure}[!htp]
\begin{center}
\advance\leftskip -3cm
\advance\rightskip -3cm
\includegraphics[width=0.65\linewidth,angle=-90]{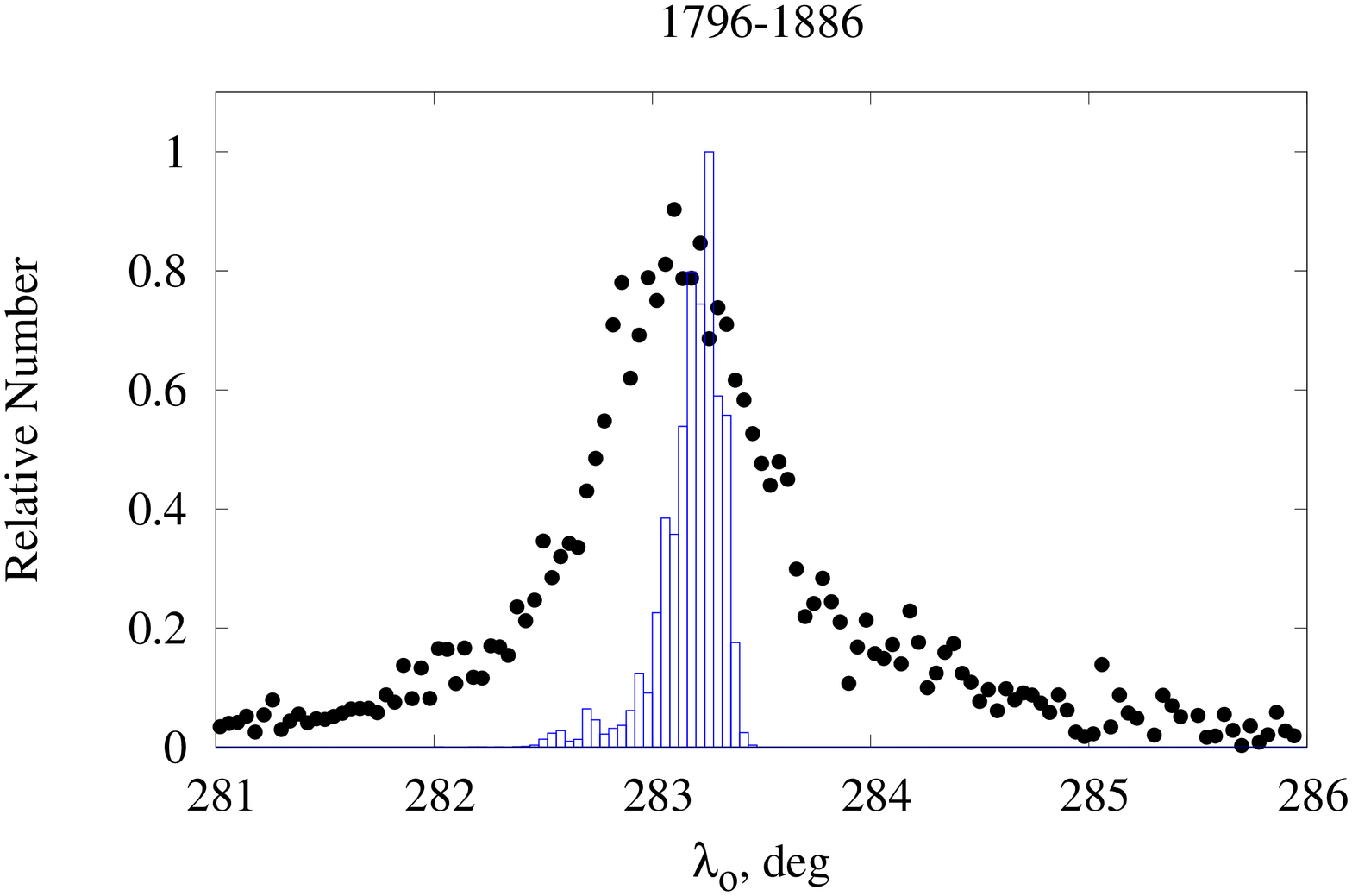}
\caption{\footnotesize{The simulated radar activity profile (blue boxes), superimposed over the observed CMOR - measured activity profile (black dots) from single-station echo observations following the same procedure described in \citep{ye14} and \citep{brown-jones95}. The theoretical profile is a stack of 15 perihelion passages of 2003 EH$_1$, corresponding to meteoroid ejections from 1796 - 1886~AD. Only meteoroids approaching the Earth's orbit within 0.01 AU are plotted. The observed activity profile is the average radar activity profile of the years 2002 to 2014.}}
\label{actprof-radar}
\end{center}
\end{figure}

\indent Fig.~\ref{actprof-radar} shows the stacked theoretical and observed radar activity profiles. The theoretical width of the core is narrower than the observed one, suggesting that there may be an older envelope of meteoroids encompassing the young core. That is, while the central portion appears to be relatively young, the wings are not.

\begin{figure}[!htp]
\begin{center}
\advance\leftskip -3cm
\advance\rightskip -3cm
\includegraphics[width=0.95\linewidth,angle=0]{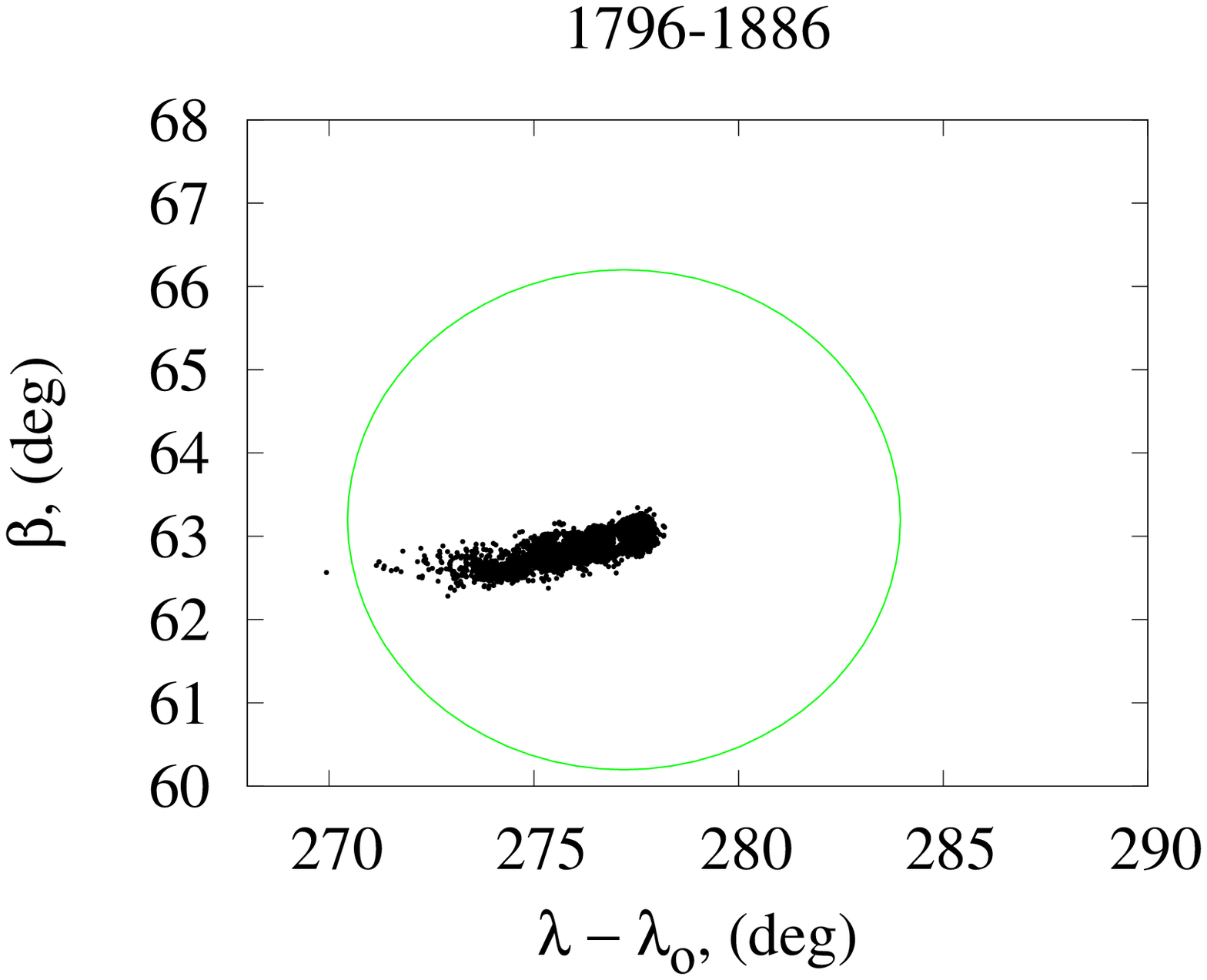}
\caption{\footnotesize{The simulated radiants position of the radar Quadrantids (black dots), along with the observed (CMOR) mean radiant position and dispersion. The simulated radiants are from $\approx$ 15 perihelion passages of 2003 EH$_1$, corresponding to meteoroid ejections from 1796 - 1886 AD. Only meteoroids approaching the Earth's orbit within 0.01 AU are plotted.}}
\label{radiant-radar}
\end{center}
\end{figure}

\indent Fig.~\ref{radiant-radar} shows the simulated distribution of radar size Quadrantids ejected between 1796 and 1886. In this case, the theoretical radiants yield an even better match to the observed one than the visual Quadrantids (Fig.~\ref{radiant-vis}). However, meteors detected by radar techniques usually have a greater radiants scatter than photographic and video techniques. Thus, the better radiant match may simply be a result of a larger observed radiant dispersion.

\indent Some authors have argued for a much older age (older than 1000 years) of the Quadrantid meteoroid stream \citep{williams79, hugh79, Williams93}. Therefore we decided to test the hypothesis whether the core of the Quadrantid meteor shower can be reproduced by meteoroid ejections around 1000 AD and 1500 AD. 

\begin{figure}[!htp]
\begin{center}
\advance\leftskip -3cm
\advance\rightskip -3cm
\includegraphics[width=0.95\linewidth,angle=0]{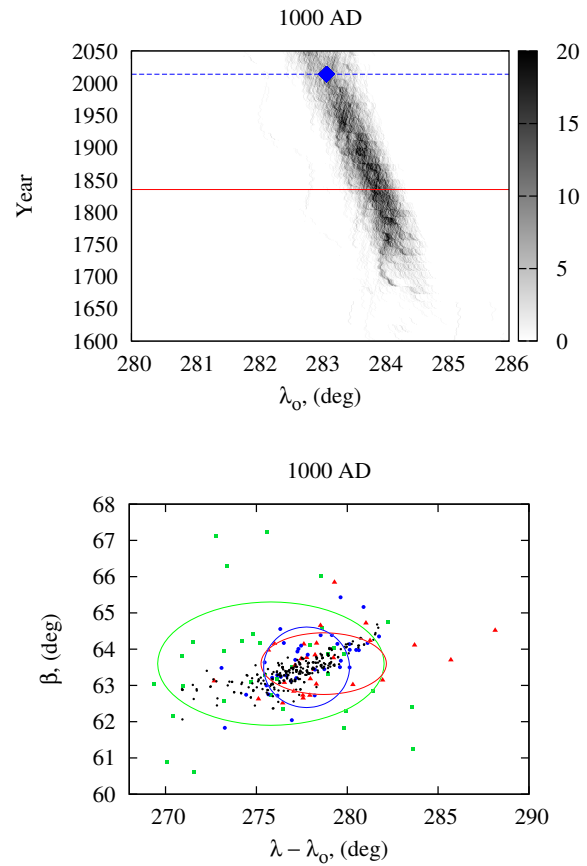}
\caption{\footnotesize{Evolution of the descending nodes of visual size meteoroids (upper panels), ejected from 2003 EH$_1$ in 1000 AD until the  present time. The red line marks the year of 1835 AD. The blue line marks the year of 2014, whereas the blue diamond corresponds to the solar longitude ($\lambda$= 283.2$^{\circ}$) of the peak of the radar shower. \\
\indent The lower panels show the simulated radiant position of meteoroids (black dots) superimposed over the observed mean radiant position, as measured by video and photographic techniques (see Fig.~\ref{vis-nd-rad}). In both figures, only simulated meteoroids approaching the Earth's orbit within 0.01 AU are plotted.}}
\label{vis-1000}
\end{center}
\end{figure}

\begin{figure}[!htp]
\begin{center}
\advance\leftskip -3cm
\advance\rightskip -3cm
\includegraphics[width=0.95\linewidth,angle=0]{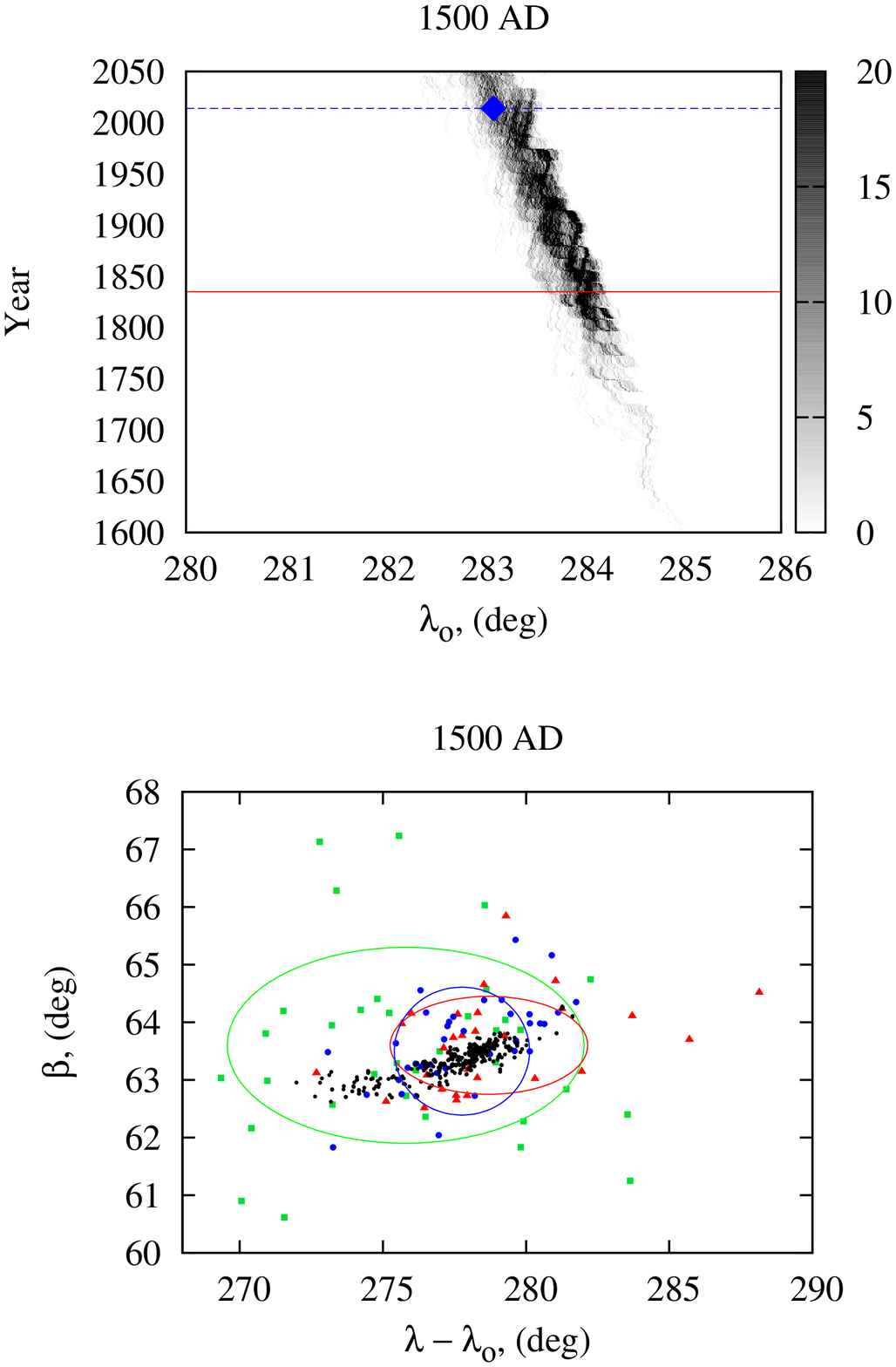}
\caption{\footnotesize{Evolution of the descending nodes of visual size meteoroids (upper panels), ejected from 2003 EH$_1$ in 1500 AD, until the present time. The red line marks the year of 1835 AD. The blue line marks the year of 2014, whereas the blue diamond corresponds to the solar longitude of the peak of the radar shower. \\
\indent The lower panels show the simulated radiant positions (black dots) superimposed over the observed mean radiant position, as measured by video and photographic techniques (see Fig.~\ref{vis-nd-rad}). In both figures, only simulated meteoroids approaching the Earth's orbit within 0.01 AU are plotted.}}
\label{vis-1500}
\end{center}
\end{figure}

\indent Fig.~\ref{vis-1000} and~\ref{vis-1500} (upper panels) show the evolution of the descending nodes of meteoroids producing visual meteoroids, ejected from 2003 EH$_1$ in 1000 AD and 1500 AD respectively. In this scenario we ejected 10$^4$ particles in a single perihelion passage of 2003 EH$_1$ in either 1000 AD and 1500 AD.

\indent Meteoroids ejected in 1000 AD and 1500 AD, show a better fit to the observed radiants positions (Fig.~\ref{vis-1000} and~\ref{vis-1500} lower panels) than ejection circa 1800 AD. However, in both cases the Quadrantid shower seems to first appear on the sky around 1650 AD - too early to be consistent with first report of the shower around 1835. Furthermore, the same procedure was applied to radar sized particles which also appeared as early as 1650 AD. We also cannot uniquely distinguish which of the observed Quadrantids may be associated with the core and which are related to the activity wings which overlap in time. 

\section{Discussion and Conclusions}

\indent We have used eight high-precision photographic Quadrantid orbits and integrated their orbits backward in time, along with the assumed parent 2003 EH$_1$, in order to constrain the most likely age of the core of the Quadrantid meteoroid stream. Out of the eight Quadrantids, five belong directly to the core of the stream having been observed within a day of the peak of the shower. In addition, five of the best quality radar Quadrantids detected by CMOR in 2013 were used as a complementary data set, in order to compare backward integrations of core radar-sized particles results with the visual counterpart of the stream.

\indent The most likely age of the core of the Quadrantid meteoroid stream, is $\approx$ 200 - 300 years, based on the backward integrations. This epoch was then used as a formation time window to test if the present observed characteristics of the radar and visual-sized particles in the 'core' of the stream could be explained by gas-drag ejection of meteoroids from 2003 EH$_1$ from this epoch. 

\indent During the backward integrations, different values (0,10$^{-5}$, 10$^{-4}$, 10$^{-3}$) for $\beta$ (solar radiation pressure to solar gravity) were considered in order to properly reflect the uncertainty in the mass of each individually observed Quadrantid. Our results did not show a noticeable difference as a function of different $\beta$ values, which is not surprising given the timescale (200-300 years) in our simulations.

\indent Our forward simulations indicate that an onset of cometary activity in 2003~EH$_1$ circa 1790 - 1796 provides a good match to all the observed constraints for the stream. In particular, the onset time of the shower in these forward simulations is near 1830, consistent with historical data. Moreover, the location of the activity peak and the width of the core of the simulated activity profile are well reproduced in both photographic and radar simulations assuming ejection from this epoch. The broader (several day duration) weak activity surrounding the shower peak is not reproduced in our simulations, leading us to conclude that while the core of the shower appears to be young, the wings are not. The simulated radiants of the radar and visual showers also match observations well. We note that the simulated visual radiant is displaced approximately 1 degree from the mean observed radiant, but this is still reasonable as it lies within the broader scatter of the overall Quadrantid radiant \citep{jen97}.

\indent In our forward modeling, for the magnitude of the meteoroid ejection speeds we used the model by \citet{BJ98}. However, in order to check the robustness of our final results against different meteoroid ejection models, we repeated our forward simulations using the meteoroid ejection model, resulting from Crifo's distributed coma model \citep{crifo95, crifo97}, which generally yields slightly lower meteoroid ejection speeds compared to other models. In spite of the latter model being more sophisticated, the differences in the final results were too small to change our conclusions. This perhaps should not be surprising given the time scales in our forward integrations (200-300 years). However, perhaps the difference in the final outcomes would be significant for relatively young meteoroid trails (as young as a few orbital periods of the parent) which requires certain meteoroid ejection conditions to be fulfilled in order for the trail to intersect the Earth's orbit.     

\indent In a summary, our forward integrations are able to reproduce the main features of the currently visible core portion of the Quadrantid stream using the formation epoch suggested from backward simulations ($\approx$ 1800 AD years). The wider portions of the shower (lasting a few days) and the recently discovered \citep{brown10} much longer radar activity (suggested as being a few months) are likely much older.

\indent At this time, it is difficult to disentangle the real nature of 2003 EH$_1$, i.e a dormant comet or an asteroid. Could 2003~EH$_1$ have been active around 1700 - 1900 AD and gone undetected? It seems quite possible.
Knowing the absolute asteroidal magnitude of 2003~EH$_1$ (H=16.2, \url{http://ssd.jpl.nasa.gov/sbdb.cgi}) we calculated its visibility from Earth throughout the 19th century. Though 2003 EH$_1$ reached almost 15th magnitude for a few brief periods in the 1800's, it could easily have escaped detection, as its average apparent magnitude was above 22. Of course, substantial cometary activity would have increased its brightness.  According to Kronk (Cometography volume 2, p ix, 2003), in the 1800's, small refractors (15-20~cm) were the most commonly used for comet observations. Larger ones were built towards the end of the century, with the 102~cm Yerkes telescope becoming active in the late century, with "...the result was that early in the century comets were usually lost after having faded to magnitude 11 or 12, while at the end of the century comets were being followed until near magnitude 16...'' \citep{kronk03}. So it is plausible that even several magnitudes of brightening could have gone unnoticed at this time. Given the current inactive state of 2003 EH$_1$ it is entirely possible it was simply too faint, even if weakly active, to be detected as a comet.

\indent What might have triggered activity in a dormant parent? The orbit itself is stable in the century preceding our proposed start time, so it was not a dramatic orbital shift. Also, during the last few hundred years the perihelion distance of 2003~EH$_1$ has been consistently increasing, making it implausible that increasing
solar heating triggered fresh activity. The ascending node of 2003 EH$_1$ was near Jupiter in the 1700-1800's, which makes collision with a main-belt asteroid unlikely, though a collision with a Jupiter Trojan remains a possibility. The descending node at the same time was near the Earth's orbit, though no close approaches between any clones and Earth were recorded in our simulations. Thus a tidal encounter with Earth is not a likely cause either.

\indent We note that there was a relatively close approach between many of the clones of 2003~EH$_1$ and Jupiter in 1794, where the minimum distance reached 0.83 Hill radii. This was the closest approach to the giant planet since 1663 when it reached 0.69 Hill radii. Thus, a reasonable explanation is that either tidal effects from the encounter with Jupiter or a collision with a Trojan asteroid activated the parent for a time, with the activity subsequently declining again to zero. However, the mechanisms of cometary activation and dormancy remain unclear and our proposed explanation is certainly not the only possibility.

\section{Acknowledgements}

This work was funded in part by the NASA Meteoroid Environment Office under NASA co-operative agreement NNX11AB76A, and partly by grant \textnumero ~P209/11/1382 from the GA{\v{C}}R, Praemium Academiae of the Czech Academy of Sciences.

\bibliographystyle{model2-names}
\bibliography{quads}







\end{document}